\documentclass[preprint,showpacs,preprintnumbers,amsmath,amssymb,11pt]{revtex4}

\usepackage{graphicx}
\usepackage{dcolumn}
\usepackage{bm}
\usepackage{psfrag}

\begin{document}

\preprint{BNL-HET-03/24}
\preprint{Oct. 2003; Jan. 2004, v2}

\vskip 1.5in

\title{One-loop Radiative Corrections to the $\rho$ Parameter\\
 in 
the Littlest Higgs Model}%
\vskip 1in
\author{Mu-Chun Chen\footnote{chen@quark.phy.bnl.gov} 
and Sally Dawson\footnote{dawson@bnl.gov}}
\affiliation{
High Energy Theory Group, Department of Physics\\
Brookhaven National Laboratory, 
Upton, NY 11973-5000, U.S.A.
}%


\vskip 2in

\begin{abstract}
We perform a one-loop analysis of the $\rho$ parameter 
in the Littlest Higgs model, including the logarithmically enhanced 
contributions from both fermion and scalar loops. We find the one-loop 
contributions are comparable to the tree level corrections
in some regions of parameter space. 
The fermion loop contribution 
dominates in the low cutoff scale $f$ region. On the other hand,  
the scalar loop contribution dominates in the high cutoff 
scale $f$ region and it grows with the cutoff scale $f$. This in turn 
implies an upper bound on the cutoff scale.   
A low cutoff scale is allowed for a non-zero triplet VEV. Constraints 
on various other parameters in the model are also discussed.
The role of triplet scalars in constructing a consistent 
renormalization scheme is emphasized.
\end{abstract}

\maketitle
\section{Introduction}

The Standard Model(SM) requires a Higgs boson to explain the generation
of fermion and gauge boson masses.  Precision electroweak measurements suggest
that the Higgs boson must be relatively light, $m_{H} <219~GeV$~\cite{lep03}. 
Currently, experimental data overwhelmingly support the SM with 
a light Higgs boson. The
simplest version of the Standard Model with a single Higgs boson,
however, has the theoretical problem that 
the Higgs boson mass is quadratically sensitive 
to any new physics which may
arise at high energy scales.  Fine tuning and naturalness
arguments suggest that the  scale at
which this new physics enters should be on the order of a TeV.  

Supersymmetry addresses the quadratic sensitivity of the SM to
high mass scales by introducing superpartners to the ordinary
fields.  The contributions of the superpartners to the Higgs mass
explicitly cancel the quadratic dependence of the Higgs mass
on the high mass scales.
Little Higgs (LH) 
models~\cite{Arkani-Hamed:2001nc,Arkani-Hamed:2002pa,Arkani-Hamed:2002qx,Arkani-Hamed:2002qy,Low:2002ws,Kaplan:2003uc,Chang:2003un,Skiba:2003yf,Chang:2003zn} 
are a new approach to understanding
the hierarchy between the $TeV$ scale of possible
new physics and the
electroweak scale, $v=246~GeV=(\sqrt{2} G_F)^{-1/2}$.
   These models have an expanded
gauge structure at the TeV scale which contains the Standard Model
$SU(2)\times U(1)$ electroweak gauge groups.  
The LH models are constructed
such that an approximate global symmetry prohibits the Higgs boson from 
obtaining a quadratically divergent mass until at least two loop order.
The Higgs boson is a pseudo-Goldstone 
boson~\cite{Dimopoulos:1982xc,Kaplan:1984fs,Kaplan:1984sm,Georgi:1984af,Georgi:1984ef,Dugan:1985hq,Banks:1984gj} 
resulting from the spontaneous breaking of the
approximate global symmetry and so
 is naturally light.  The Standard Model then emerges
as an effective theory which is valid below the scale $f$
associated with the spontaneous breaking of the global symmetry.

Little Higgs models  contain weakly coupled TeV scale gauge bosons 
from the expanded gauge structure, which
couple to the Standard Model fermions.  In addition, these
new gauge bosons typically mix with the Standard Model $W$ and
$Z$ gauge bosons.  Modifications of the electroweak
sector of the theory, however, are severely restricted by precision
electroweak data and  require  the scale of the little Higgs
physics, $f$, to be in the range 
$f>1-6~TeV$~\cite{Csaki:2002qg,Hewett:2002px,Csaki:2003si,Kribs:2003yu,Gregoire:2003kr,Perelstein:2003wd,Casalbuoni:2003ft,Kilian:2003xt}, 
depending on the specifics 
of the model.  The LH models also contain expanded Higgs sectors with
additional Higgs doublets and triplets, as well as a new charge $2/3$
quark, which have importance implications for precision electroweak
measurements.  

In this paper, we analyze the contributions of the heavy fermions and
scalars to the isospin violating $\rho$ parameter.  We include
the logarithmically enhanced  loop
corrections due to the scalar triplet which is present in such models.  
In Section II, we review the LH models.  Section III contains a 
description of our calculation, while numerical results are 
presented in Section IV.  Details of the calculation are relegated
to the appendices.

\section{Basics of Little Higgs Models} 

The Little Higgs model has been described in detail
elsewhere, but we include a brief
description of the model here
in order to clarify our notation. [Our discussion follows Ref. \cite{Han:2003wu}.]  
The minimal version, the  "littlest Higgs model" (LLH)  
\cite{Arkani-Hamed:2002qy} is a non-linear sigma model based 
on an SU(5) global symmetry, which contains a gauged 
$[SU(2) \otimes U(1)]_{1} \otimes [SU(2) \otimes U(1)]_{2}$ 
symmetry as its subgroup.  We concentrate on this model here,
although many alternatives have been 
proposed~\cite{Low:2002ws,Kaplan:2003uc,Chang:2003un,Skiba:2003yf,Chang:2003zn}.

The global SU(5) symmetry of the LLH model is broken down to SO(5) by 
the vacuum expectation value (VEV) of a sigma field~\cite{Arkani-Hamed:2002qy},
\begin{equation}
\Sigma_{0} = \left(
\begin{array}{ccc}
& & \mathbb{I}\\
& 1 &\\
\mathbb{I} & &
\end{array}\right) \quad ,
\end{equation}
where $\mathbb{I}$ is a $2 \times 2$ identity matrix
and $\langle \Sigma_0\rangle \sim f$. In addition, 
the VEV of the sigma field breaks the gauged symmetry 
$[SU(2) \otimes U(1)]_{1} \otimes [SU(2) \otimes U(1)]_{2}$ 
to its diagonal subgroup, $SU(2) \times U(1)$, which is then identified as the 
SM gauge group. The breaking of the global symmetry, $SU(5) \rightarrow SO(5)$, 
leaves $14$ Goldstone bosons, $\Pi \equiv \pi^{a} X^{a}$, 
which can be written as
\begin{equation}
\Sigma = e^{i\Pi/f} \Sigma_{0} e^{i\Pi^{T}/f} 
= \Sigma_{0} + \frac{2i}{f} \Pi \Sigma_{0} + \mathcal{O}(1/f^{2}),
\end{equation}
where $X^a$ correspond to the broken $SU(5)$ generators.
Four of these Goldstone bosons become the longitudinal components
of the broken gauge symmetry, while
the remaining ten pseudo-Goldstone bosons can be parameterized 
as~\cite{Arkani-Hamed:2002qy},
\begin{equation}
\Pi = \left(\begin{array}{ccc}
 & h^{\dagger}/\sqrt{2} & \Phi^{\dagger}\\
h/\sqrt{2} & & h^{\ast}/\sqrt{2}\\
\Phi & h^{T}/\sqrt{2} &
\end{array}\right),
\end{equation}
where $h$ is identified as the SM Higgs doublet, $h=(h^{+}, \; h^{0})$, and 
$\Phi$ is a complex
 SU(2) triplet with hypercharge $Y=2$, 
\begin{equation}
\Phi = \left(\begin{array}{cc}
\Phi^{++} & \Phi^{+}/\sqrt{2} \\
\Phi^{+}/\sqrt{2} & \Phi^{0}
\end{array}\right).
\end{equation}
The existence of an $SU(2)$ triplet is a general
feature of models of this type.


The Lagrangian is given by
\begin{equation}
\mathcal{L} = \mathcal{L}_{k} + \mathcal{L}_{Yuk} \quad ,
\end{equation}
where $\mathcal{L}_{k}$ contains the kinetic terms of all fields and 
$\mathcal{L}_{Yuk}$ describes the Yukawa interactions. The gauge bosons 
acquire their masses through the kinetic terms of the $\Sigma$ field
\begin{equation}\label{sigmakin}
\mathcal{L}_{k}= \frac{f^{2}}{8} \mbox{Tr} \{ (D_{\mu} \Sigma) 
\, (D^{\mu} \Sigma)^{\dagger} \} \quad ,
\end{equation}
where the covariant derivative of the $\Sigma$ field is defined as
\begin{equation}
D_{\mu} \Sigma = \partial_{\mu} \Sigma 
- i \sum_{j} [ \, g_{j} W_{j}^{a} (Q_{j}^{a}\Sigma + \Sigma Q_{j}^{a}{}^{T})
+ g_{j}^{'} B_{j} (Y_{j} \Sigma+\Sigma Y_{j}^{T}) \, ] \quad .
\end{equation}
The SU(2) gauge fields are given by $W_{j} = \sum_{a=1}^{3} \; W_{j}^{\mu \, a} 
Q_{j}^{a}$, and the U(1) gauge fields are $B_{j} = B_{j}^{\mu} Y_{j}$,
with gauge couplings $g_1,g_2$ and $g_1^\prime,g_2^\prime$.
(The $SU(2)$ and hypercharge, $Y_{j}$ assignments can be found in 
Ref.~\cite{Arkani-Hamed:2002qy}).  

The $\Sigma_0$ VEV generates masses and mixing between the gauge bosons.
The heavy gauge boson mass eigenstates are given by,
\begin{equation}
W_{H}^{a} = -c W_{1}^{a} + s W_{2}^{a}, \quad 
B_{H}^{a} = -c^{\prime} B_{1} + s^{\prime} B_{2} \quad ,
\end{equation}
with masses~\cite{Csaki:2002qg,Hewett:2002px,Han:2003wu} 
\begin{equation}
M_{\scriptstyle W_{H}}^{2} = \frac{f^{2}}{4}(g_{1}^{2}+g_{2}^{2}), \quad
M_{\scriptstyle B_{H}}^{2} = \frac{f^{2}}{20} (g_{1}^{'2}+g_{2}^{'2}) \quad .
\end{equation}
The orthogonal combinations of gauge bosons are identified as the SM
$W$ and $B$, with couplings~\cite{Csaki:2002qg,Hewett:2002px,Han:2003wu},
\begin{equation}
g = g_{1}s = g_{2}c, \quad 
g^{\prime} = g_{1}^{\prime}s^{\prime} = g_{2}^{\prime}c^{\prime} \quad.
\end{equation} 
The mixing between the two SU(2)'s (U(1)'s) is described by the parameters 
$s$  and $s^{\prime}$~\cite{Csaki:2002qg,Hewett:2002px,Han:2003wu},
\begin{equation}
s = \frac{g_{2}}{\sqrt{g_{1}^{2}+g_{2}^{2}}}, \quad
s^{\prime} = \frac{g_{2}^{\prime}}{\sqrt{g_{1}^{\prime 2}+g_{2}^{\prime 2}}} 
\quad .
\end{equation}
(and $c=\sqrt{1-s^{2}}$, $c^{\prime} = \sqrt{1-s^{\prime 2}}$.)
The coupling of fermions to the photon is then given 
by~\cite{Csaki:2002qg,Hewett:2002px,Han:2003wu},
\begin{eqnarray}
e & = & gs_{W}
\label{eggp}\\
s_{W} & = & \frac{g^{\prime}}{\sqrt{g^{2} + g^{\prime 2}}}
\label{eggp1}\\
c_{W} & = & \frac{g}{\sqrt{g^{2} + g^{\prime 2}}} \quad .
\label{eggp2}
\end{eqnarray}

In the Yukawa sector, a new vector-like charge $2/3$ fermion
is introduced to cancel the quadratic sensitivity of the Higgs
mass to the top quark loops.  This cancellation fixes the
Yukawa interactions~\cite{Arkani-Hamed:2002qy,Han:2003wu},
\begin{equation}
\mathcal{L}_{Yuk} = {1\over 2}
\lambda_{1} f \epsilon_{ijk} \epsilon_{xy} \chi_{i} 
\Sigma_{jx}\Sigma_{ky} u_{3}^{\prime}{}^{c} 
+ \lambda_{2} f \tilde{t} \tilde{t}^{c} + h.c. \quad ,
\end{equation}
where $t_3$ is the SM top quark, $u_{3}^\prime$ is the SM right-handed top quark, 
$({\tilde t},\tilde{t}^{\prime c})$ is a new charge $2/3$
vector-like quark and ${\chi}=(b_{3}, \; t_{3}, \; \tilde{t})$.
Expanding the $\Sigma$ field in terms of its component fields, 
the mass terms of the fermions 
are~\cite{Arkani-Hamed:2002qy,Hewett:2002px,Han:2003wu},
\begin{eqnarray}
\mathcal{L}_{Yuk} & = & 
f \biggl[ \sqrt{\lambda_{1}^{2}+\lambda_{2}^{2}} 
- \frac{\lambda_{1}^{2}}{2\sqrt{\lambda_{1}^{2}+\lambda_{2}^{2}}}
\frac{v^{2}}{f^{2}} \biggr] \tilde{t}\tilde{t}^{c} 
- \frac{i\lambda_{1}^{2}v}{\sqrt{\lambda_{1}^{2}+\lambda_{2}^{2}}} t_{3} \tilde{t}^{c}
\\
&&
- \frac{i\lambda_{1}\lambda_{2}v}{\sqrt{\lambda_{1}^{2}+\lambda_{2}^{2}}}
t_{3}u_{3}^{c} 
- \frac{\lambda_{1}\lambda_{2}}{2\sqrt{\lambda_{1}^{2}+\lambda_{2}^{2}}} 
\frac{v^{2}}{f^{2}} 
\tilde{t}u_{3}^{c} \quad .
\nonumber
\end{eqnarray}  
The following mass eigenstates are obtained after diagonalizing the above 
mass terms~\cite{Hewett:2002px,Han:2003wu},
\begin{eqnarray}
t_{L} & = & t_{3} + i x_{L} \frac{v}{f} \tilde{t}\\
T_{L} & = & \tilde{t} -i x_{L} \frac{v}{f} t_{3}\\
t_{R}^{c} & = & u_{3}^{c} = \frac{1}{\sqrt{\lambda_{1}^{2}+\lambda_{2}^{2}}}
(-\lambda_{1}\tilde{t}^{\, 'c} + \lambda_{2} u_{3}^{'c})\\
T_{R}^{c} & = & \tilde{t}^{c} = \frac{1}{\sqrt{\lambda_{1}^{2}+\lambda_{2}^{2}}}
(\lambda_{2}\tilde{t}^{\, 'c} + \lambda_{1} u_{3}^{'c}) \quad .
\end{eqnarray} 
We express our results in terms of $x_{L}$,
which  parameterizes the mixing between $t_{3}$ and $\tilde{t}$; it is 
given by~\cite{Hewett:2002px,Han:2003wu}
\begin{equation}
x_{L} =  \frac{\lambda_{1}^{2}} 
{\lambda_{1}^{2}+\lambda_{2}^{2}} \quad .
\end{equation}
The tree level $t\overline{t}h$ Yukawa coupling is now~\cite{Hewett:2002px,Han:2003wu},
\begin{equation}
y_{t} \, = \,  {m_{t} \over  v} \, = \, 
\frac{i \lambda_{1} \lambda_{2}}{\sqrt{\lambda_{1}^{2}+\lambda_{2}^{2}}} \, 
\biggl[ \, 1 + \frac{v^{2}}{2f^{2}} 
\frac{\lambda_{1}^{2}}{\lambda_{1}^{2}+\lambda_{2}^{2}}
(1+\frac{\lambda_{1}^{2}}{\lambda_{1}^{2}+\lambda_{2}^{2}}) \, \biggr].
\end{equation}
In  the limit that the cut-off scale $f$ goes to infinity, 
the coupling~\cite{Hewett:2002px,Han:2003wu}
\begin{equation}
y_{t} \, = \, 
\frac{i \lambda_{1} \lambda_{2}}{\sqrt{\lambda_{1}^{2}+\lambda_{2}^{2}}}
\end{equation}
is identified as the top quark Yukawa coupling of the SM.

The one-loop quadratically divergent contributions to the Coleman-Weinberg 
potential due to the scalars and fermions are given 
by~\cite{Arkani-Hamed:2002qy,Han:2003wu}
\begin{eqnarray}
\mathcal{L}_{s} & = & \frac{a}{2} f^{4} \{ g_{j}^{2} 
\sum_{a} \; Tr [ ( Q_{j}^{a} \Sigma ) +
( Q_{j}^{a} \Sigma) {}^{\ast} ]
+ g_{j}^{\prime 2} 
Tr \; [ ( Y_{j} \Sigma ) + 
( Y_{j} \Sigma ){}^{\ast} ] \}
\label{cwps}\\
\mathcal{L}_{f} & = & -\frac{a^{\prime}}{4} \lambda_{1}^{2} 
f^{4} \epsilon^{wx}\epsilon_{yz} \epsilon^{ijk} \epsilon_{kmn} 
\Sigma_{iw} \Sigma_{jx} \Sigma^{\ast my} \Sigma^{\ast nz} \quad .
\label{cwpf}
\end{eqnarray}
where $a$ and $a^\prime$ are unknown coefficients parameterizing 
physics from the Ultra-Violet (UV) completion. 
These lead to the following Coleman-Weinberg 
potential~\cite{Arkani-Hamed:2002qy,Han:2003wu} 
\begin{equation}
V_{CW} = \lambda_{\Phi^{2}} f^{2} \; Tr (\Phi^{\dagger}\Phi) 
+ i \lambda_{h\Phi h} f (h \Phi^{\dagger} h^{T}-h^{\ast} \Phi h^{\dagger})
-\mu^{2} hh^{\dagger} + \lambda_{h^{4}} (hh^{\dagger})^{2} \quad ,
\end{equation}
where~\cite{Arkani-Hamed:2002qy,Han:2003wu} 
\begin{eqnarray}
\mu^{2} & \sim & \frac{f^{2}}{16\pi^{2}}
\label{eq:mu2}
\\
4 \lambda_{h^{4}} & = & \lambda_{\Phi^{2}} = 
\frac{a}{2} [ \frac{g^{2}}{s^{2}c^{2}} 
+ \frac{g^{\prime 2}}{s^{\prime 2}c^{\prime 2}} ] + 8a^{'}\lambda_{1}^{2}
\\
\lambda_{h \Phi h} & = & 
-\frac{a}{4} [ \frac{g^{2}(c^{2}-s^{2})}{s^{2}c^{2}} 
+ \frac{g^{\prime 2}(c^{\prime 2}-s^{\prime 2})}{s^{\prime 2}c^{\prime 2}} 
] + 4a^{'}\lambda_{1}^{2} 
\quad .
\end{eqnarray}
and $\mu^{2}$ is generated by one-loop logarithmically divergent and two-loop 
quadratic divergent contributions. 
The VEV's of the SM Higgs doublet and the SU(2) triplet 
are 
$\left< h^{0} \right> = v/\sqrt{2}$ 
and  $\left< \Phi^{0} \right> = -i v^{\prime}$, 
where~\cite{Arkani-Hamed:2002qy,Han:2003wu}
\begin{eqnarray}
v^{2} & = & \frac{\mu^{2}}{\lambda_{h^{4}}
-\frac{\lambda_{h\Phi h}^{2}}{\lambda_{\Phi^{2}}}}
\\
v^{\prime} & = & \frac{\lambda_{h\Phi h}}{2\lambda_{\Phi^{2}}} \frac{v^{2}}{f}
\end{eqnarray}
To obtain the correct electroweak symmetry breaking vacuum with $m_{H}^{2} > 0$ 
and $M_{\Phi}^{2} > 0$, the following conditions must be 
satisfied~\cite{Arkani-Hamed:2002qy,Han:2003wu},
\begin{eqnarray}
\lambda_{h^{4}} - \frac{\lambda_{h\Phi h}^{2}}{\lambda_{\Phi^{2}}} > 0
\\
\frac{v^{'}}{v} < \frac{1}{4} \frac{v}{f}
\label{vvprelation}
\quad .
\end{eqnarray}

We summarize in Table \ref{mass} the mass spectrum 
of the model~\cite{Han:2003wu} and in Tables II, III, and 
IV~\cite{Han:2003wu} the relevant couplings.
\begin{table}
\caption{\label{mass} Mass spectrum of the gauge bosons, scalar fields and 
the fermions. 
The parameters $m_{W}$ and $m_{Z}$ are given by the SM expressions, $m_{W}=gv/2$ 
and $m_{Z}=gv/2c_{W}$, respectively, where $c_{W}$ is defined in 
Eq.(\ref{eggp})-(\ref{eggp2}). 
The parameter $x_{H}$ is a mixing parameter  
in the neutral gauge boson sector, $x_{H} = [5gg^{'}scs^{'}c^{'}(c^{2}s^{'2}+
s^{2}c^{'2})]/[2(5g^{2}s^{'2}c^{'2}-g^{'2}s^{2}c^{2})]$~\cite{Han:2003wu}.}
\begin{tabular}{|c| c l|}
\toprule
gauge boson $x$ & \hspace{1cm} & $M_{x}^{2}$
\\
\hline
$W_{L}^{\pm}$ & & $m_{W}^{2} [ 1 - 
\frac{v^{2}}{f^{2}}
(\frac{1}{6} + \frac{1}{4}(c^{2}-s^{2})^{2})+4\frac{v^{'2}}{v^{2}}]$
\\
$W_{H}^{\pm}$ & & $m_{W}^{2}[\frac{f^{2}}{c^{2}s^{2}v^{2}}-1]$
\\
$A_{L}$ & & $0$
\\
$Z_{L}$ & & $m_{Z}^{2}[1-\frac{v^{2}}{f^{2}}(\frac{1}{6} 
+ \frac{1}{4}(c^{2}-s^{2})^{2} + \frac{5}{4}(c^{'2}-s^{'2})^{2})
+ \frac{8v^{'2}}{v^{2}}]$
\\
$A_{H}$ & & $m_{Z}^{2}s_{W}^{2}[\frac{f^{2}}{5s^{'2}c^{'2}v^{2}}-1
+\frac{x_{H}c_{W}^{2}}{4s^{2}c^{2}s_{W}^{2}}]$
\\
$Z_{H}$ & & $m_{W}^{2}[\frac{f^{2}}{s^{2}c^{2}v^{2}}-1
-\frac{x_{H}s_{W}^{2}}{s^{'2}c^{'2}c_{W}^{2}}]$
\\
\hline
scalar field $s$ & & $M_{s}^{2}$
\\
\hline
$h$ & & $2\mu^{2} = 2(\lambda_{h^{4}}-\lambda_{h\Phi h}^{2}/\lambda_{\Phi^{2}}) v^{2}  
\equiv m_{H}^{2}$ 
\\
$\Phi$ & & 
$\lambda_{\Phi^{2}} f^{2} = 
2 m_{H}^{2} \frac{f^{2}}{v^{2}} \frac{1}{1-(\frac{4v^{'}f}{v^{2}})^{2}}$
\\
\hline
fermion $f$ & & $m_{f}$
\\
\hline
$t$ & & $\frac{\lambda_{1}\lambda_{2}}{\sqrt{\lambda_{1}^{2}+\lambda_{2}^{2}}} v
[1 + \frac{v^{2}}{2f^{2}}\frac{\lambda_{1}^{2}}{\lambda_{1}^{2}+\lambda_{2}^{2}}
(1+\frac{\lambda_{1}^{2}}{\lambda_{1}^{2}+\lambda_{2}^{2}})]$
\\
$T$ & & 
$\sqrt{\lambda_{1}^{2}+\lambda_{2}^{2}}f 
[1 - \frac{v^{2}}{2f^{2}}\frac{\lambda_{1}^{2}}{\lambda_{1}^{2}+\lambda_{2}^{2}}
(1+\frac{\lambda_{1}^{2}}{\lambda_{1}^{2}+\lambda_{2}^{2}})]$
\\
\botrule
\end{tabular}
\end{table}

\section{ The Renormalization Procedure}

Precision electroweak measurements give stringent bounds on the scale of
little Higgs type 
models~\cite{Csaki:2002qg,Hewett:2002px,Csaki:2003si,Kribs:2003yu,Gregoire:2003kr,Perelstein:2003wd,Casalbuoni:2003ft,Kilian:2003xt}.  
One of the strongest bounds comes from fits
to the $\rho$ parameter, since in
the LLH model the relation $\rho=1$ is modified  at the tree level.
While the Standard Model requires three input parameters
in the weak sector (corresponding to
the $SU(2)\times U(1)$ gauge coupling constants and the Higgs doublet
VEV, $v$), a model with $\rho\ne 1$ at tree level, such as the LLH model
or any model
with a Higgs triplet, requires an additional input parameter
in the gauge-fermion sector, which
can be taken to be the VEV of the Higgs triplet, $v^\prime$.  
The need for this additional input parameter when  $\rho\ne 1$
 at the tree level was first noted in 
Refs.~ \cite{Passarino:1990xx,Lynn:1990zk}.
This extra input parameter, beyond the three of the
Standard Model, has important implications when models with Higgs
triplets are studied beyond tree 
level\cite{Passarino:1990xx,Lynn:1990zk,Blank:1998qa,Czakon:1999ue,Czakon:1999ha}.  Many of the familiar predictions of the
Standard Model are drastically changed by the need for an extra
input parameter.  For example, the dependence of
the $\rho$ parameter on the top quark mass becomes logarithmic
(instead of quadratic as it is in the Standard Model) 
in theories with a Higgs triplet, as emphasized in Refs.
\cite{Blank:1998qa,Czakon:1999ue,Czakon:1999ha}

We choose as our input parameters
the muon decay constant $G_{\mu}$, the physical Z-boson mass $M_{Z}^{2}$, 
the effective lepton mixing angle $s_{\theta}^{2}$ and the fine-structure 
constant $\alpha(M_{Z}^{2})$ as the four independent input parameters in 
the renormalization procedure. The $\rho$ parameter, defined as,
\begin{equation}\label{eq:rhodef}
\rho \equiv \frac{M_{W_{L}}^{2}}{M_{Z}^{2}c_{\theta}^{2}},
\end{equation}
and the $W$-boson mass are then derived quantities (in contrast to
the Standard Model).
The effective leptonic mixing angle 
$s_{\theta}^{2}$ at the Z-resonance is defined as the ratio of the electron 
vector to axial vector coupling constants to the Z-boson, 
\begin{equation}\label{eq:elm}
\frac{Re(g_{V}^{e})}{Re(g_{A}^{e})} = 4s_{\theta}^{2}-1.
\end{equation}
where we have defined the coupling of a fermion $\psi_{i}$, with mass
$m_{i}$, to gauge boson $X$ as,
\begin{equation}
\mathcal{L} = i\overline{\psi}_{i} \gamma_{\mu} (g_{V} + g_{A} \gamma_{5}) 
\psi_{j} X^{\mu}\quad .
\end{equation}

The effective Lagrangian of the charged current interaction in the LLH model is
given by~\cite{Csaki:2002qg,Hewett:2002px,Han:2003wu},
\begin{eqnarray}
\mathcal{L}_{cc} & = & g W_{L_{\mu}}^{a}J^{a\mu}
(1+\frac{c^{2}(s^{2}-c^{2})v^{2}}{f^{2}})
+g^{\prime}B_{L_{\mu}}J_{Y}^{\mu}
(1-5\frac{c^{\prime 2}(s^{\prime 4}-c^{\prime 4})v^{2}}{f^{2}})
\\
& & \quad + g W_{L_{\mu}}^{3} J^{\mu}_{Y} 
\frac{5(s^{\prime 4}-c^{\prime 4})v^{2}}{f^{2}}
- g^{\prime}B_{L_{\mu}}J^{3\mu} \frac{c^{2}(s^{2}-c^{2})v^{2}}{f^{2}}
\nonumber\\
& & \quad -J_{\mu}^{a}J^{a\mu} \frac{2c^{4}}{f^{2}}
-J_{\mu}^{Y}J^{Y\mu} \frac{10c^{\prime 4}}{f^{2}}  \; .
\nonumber
\end{eqnarray}
After integrating out the W-boson, $W_{L}$, we obtain the 
muon decay constant, $G_{\mu}$, given 
by~\cite{Hewett:2002px,Csaki:2003si,Kribs:2003yu}
\begin{equation}
G_{\mu} = \frac{1}{\sqrt{2}}\{\frac{g^{2}}{4M_{W_{L}}^{2}}
[1+\frac{c^{2}(s^{2}-c^{2})v^{2}}{f^{2}}]+\frac{c^{4}}{f^{2}}\} \quad .
\end{equation} 
Replacing the W-boson mass $M_{W_{L}}^{2}$ by
\begin{equation}
M_{W_{L}}^{2} =  m_{W}^{2} 
\biggl[ 1 - \frac{v^{2}}{f^{2}}\biggl(\frac{1}{6} 
+ \frac{1}{4}(c^{2}-s^{2})^{2}\biggr)+4\frac{v^{'2}}{v^{2}}\biggr] \quad,
\end{equation}
where $m_{W}^{2}$ is given by the SM expression,
$ m_{W}= gv/2$, 
the muon decay constant $G_{\mu}$ can be written as 
\begin{equation}\label{vvg}
\frac{1}{\sqrt{2}G_{\mu}} = v^{2} \biggl(
1 + \frac{v^{2}}{4f^{2}} + 4 \frac{v^{\prime 2}}{f^{2}} \biggl) \quad ,
\end{equation} 
which is then inverted to give $v^{2}$ in terms of  
$G_{\mu}$, $f$ and $v^{\prime}$,
\begin{equation}
v^{2} = \frac{1}{\sqrt{2}G_{\mu}} \biggl[ 1 - \frac{1}{4\sqrt{2}G_{\mu}f^{2}} 
- 4 \frac{v^{\prime 2}}{f^{2}}\biggr] \; .
\end{equation}

In the LLH model, the vector and the axial vector
parts of the neutral current $Ze\bar{e}$ coupling constant 
are given by~\cite{Han:2003wu}
\begin{eqnarray}
g_{V}^{e} & = & \frac{g}{2c_{W}} \{
(-1/2 + 2s_{W}^{2}) 
\\
&& \qquad + \frac{v^{2}}{f^{2}}[-c_{W}x_{Z}^{W\prime}\frac{c}{2s}
+\frac{s_{W}x_{Z}^{B\prime}}{s^{\prime}c^{\prime}}(2y_{e}-\frac{9}{5}
+ \frac{3}{2}c^{\prime 2}) ] \} 
\nonumber\\
g_{A}^{e} & = & \frac{g}{2c_{W}} \{
\frac{1}{2} + \frac{v^{2}}{f^{2}} [ c_{W}x_{Z}^{W\prime}\frac{c}{2s}
+ \frac{s_{W}x_{Z}^{B\prime}}{s^{\prime}c^{\prime}}(-\frac{1}{5}
+\frac{1}{2}c^{\prime 2})] \}.
\end{eqnarray}
where $x_{Z}^{B\prime}$ and $x_{Z}^{W\prime}$ are given by,
\begin{eqnarray}
x_{Z}^{B\prime} & = &
-\frac{5}{2s_{W}} s^{\prime} c^{\prime} (c^{\prime2}-s^{\prime2}) \; ,
\\
x_{Z}^{W\prime} & = &
-\frac{1}{2c_{W}} sc(c^{2}-s^{2}) \; .
\end{eqnarray}
The ratio $Re(g_{V}^{e})/Re(g_{A}^{e})$ is thus given by
\begin{eqnarray} 
\frac{Re(g_{V}^{e})}{Re(g_{A}^{e})} & \equiv & 4s_{\theta}^{2} - 1
\\
& = & 
(4s_{W}^{2}-1) + \frac{2v^{2}}{f^{2}} 
[s_{W}^{2}c^{2}(c^{2}-s^{2}) - 
c_{W}^{2}(c^{\prime2}-s^{\prime 2})(-2+5c^{\prime 2})].
\nonumber
\end{eqnarray}
The effective leptonic mixing angle $s_{\theta}^{2}$ and the mixing angle 
$s_{W}^{2}$ in the LLH model are then related via the 
following relation,
\begin{equation}\label{elm}
s_{\theta}^{2} = s_{W}^{2} + \frac{v^{2}}{2f^{2}} 
[s_{W}^{2}c^{2}(c^{2}-s^{2}) - 
c_{W}^{2}(c^{\prime2}-s^{\prime 2})(-2+5c^{\prime 2})].
\end{equation}
This equation can then be inverted and gives
\begin{equation}
s_{W}^{2} = s_{\theta}^{2}+\Delta s_{\theta}^{2}
\end{equation}
where
\begin{equation}
\Delta s_{\theta}^{2}= - \frac{1}{2\sqrt{2}G_{\mu}f^{2}} 
[s_{\theta}^{2} c^{2} (c^{2}-s^{2}) - 
c_{\theta}^{2} (c^{\prime2} - s^{\prime 2})(-2+5c^{\prime 2})] \quad .
\end{equation}

The $SU(2)_{L}$ gauge coupling constant, $g$, can be re-written in terms 
of the effective leptonic mixing angle, $s_{\theta}^{2}$, and the 
fine-structure constant, $\alpha$, as
\begin{equation}
g^{2} = \frac{e^{2}}{s_{W}^{2}} = \frac{4\pi\alpha}{s_{\theta}^{2}}
(1 - \frac{\Delta s_{\theta}^{2}}{s_{\theta}^{2}}) \quad .
\end{equation}  
We then arrive at
\begin{equation}\label{gmu}
\sqrt{2} G_{\mu}= \frac{\pi \alpha}{M_{Z}^{2}s_{\theta}^{2}c_{\theta}^{2}\rho}
\biggl[ 1-\frac{\Delta s_{\theta}^{2}}{s_{\theta}^{2}}
+\frac{c^{2}(s^{2}-c^{2})}{\sqrt{2}G_{\mu}f^{2}}\biggr] + \frac{c^{4}}{f^{2}}
\quad ,
\end{equation}
where $M_{Z}^{2}$ is the physical Z-boson mass, 
\begin{eqnarray}\label{mz}
M_{Z}^{2} & = &  
\frac{\pi \alpha}{\sqrt{2}G_{\mu}s_{\theta}^{2}c_{\theta}^{2}} \biggl[
 1 - 4 \frac{v^{\prime 2}}{f^{2}} 
- \biggl( \frac{c_{\theta}^{4}
-s_{\theta}^{4}}{s_{\theta}^{2}c_{\theta}^{2}} \biggr)
\Delta s_{\theta}^{2}
\nonumber\\
& & \quad 
- \frac{1}{\sqrt{2}G_{\mu}f^{2}} 
\biggl( \frac{5}{12} + \frac{1}{4} (c^{2}-s^{2})^{2} 
+ \frac{5}{4} (c^{\prime 2} -s^{\prime 2})^{2} \biggr) 
+ 8\sqrt{2}G_{\mu}v^{\prime 2} \biggr] \quad . 
\end{eqnarray}
The left-hand side of Eq. \ref{mz} is the physical $Z$ boson mass, 
$91.1876~GeV$, while the leading contribution to the right-hand
side is $
\frac{\pi \alpha}{\sqrt{2}G_{\mu}s_{\theta}^{2}c_{\theta}^{2}}=91.475~GeV$.  
In order to obtain the correct $Z$ mass,  the sub-leading terms on
the right-hand side must be non-zero. As $f$ becomes larger, 
the tree- level corrections become smaller 
and insufficient to satisfy Eq.~\ref{mz}.

Using 
Eq.  
(\ref{eq:rhodef}), (\ref{gmu}) and (\ref{mz}), 
the parameters $M_{W_{L}}^{2}$, $\rho$ and $s$ can be derived, in terms 
of $G_{\mu}$, $M_{Z}^{2}$, $\alpha(M_{Z})$ and $s^{2}_\theta$, 
and the free parameters, $f$, $v^{\prime}$ and $s^{\prime}$. 
The $\rho$ parameter at tree level is 
\begin{equation}\label{rhotree}
\rho^{\mbox{\tiny tree}} = 
\frac{\pi \alpha}{\sqrt{2}M_{Z}^{2}s_{\theta}^{2}c_{\theta}^{2} G_{\mu}}
\biggl[ 1-\frac{\Delta s_{\theta}^{2}}{s_{\theta}^{2}}
+\frac{c^{2}s^{2}}{\sqrt{2}G_{\mu}f^{2}}
\biggr] \; ,
\end{equation}
where the parameter $s^{2}$ and $c^{2} = 1-s^{2}$ are determined 
by Eq.(\ref{mz}). Note that $\rho^{\mbox{\tiny tree}}$ depends on 
$v^{\prime}$ implicitly through $s$. 
Given the value of $\rho^{\mbox{\tiny tree}}$ in Eq.(\ref{rhotree}), 
the W-boson mass $M_{W_{L}}$ at tree 
level is determined by Eq.(\ref{eq:rhodef}).

Since the loop factor occurring in radiative corrections,  
$1/16\pi^2$, is similar in magnitude to
the expansion parameter, $v^{2}/f^{2}$, of chiral perturbation 
theory, the one-loop radiative corrections can
be comparable in size to  the next-to-leading order contributions at tree 
level of Eq.~\ref{rhotree}.
In this paper, 
we compute the loop corrections to the $\rho$ parameter which are
enhanced by large logarithms; we focus on terms of ${\cal O}\biggl(
{1\over
16\pi^2}\ln\biggl({M^2 \over Q^2}\biggr)\biggr)$, where 
$Q\sim M_Z$ and $M\sim f \sim 
{\cal O}(TeV)$.
At the one-loop level, we have to take into account the radiative correction 
to the muon decay constant $G_{\mu}$, the counterterm for the electric 
charge $e$, the mass counterterm of the Z-boson, and the counterterm 
for the leptonic mixing angle $s_{\theta}^{2}$. These corrections are collected  
in the quantity $\Delta r_{Z}$, and Eq.(\ref{gmu}) can then be rewritten 
in the following way,   
\begin{equation}
s_{\theta}^{2}c_{\theta}^{2} = 
\frac{\pi \alpha(M_{Z}^{2})}{\sqrt{2}G_{\mu}M_{Z}^{2}\rho}
\biggl[1-\frac{\Delta s_{\theta}^{2}}{s_{\theta}^{2}}
+\frac{c^{2}s^{2}}{\sqrt{2}G_{\mu}f^{2}}
+\Delta r_{Z} \biggr],
\end{equation}
where
\begin{equation}\label{deltarz}
\Delta r_{Z} = -\frac{\delta G_{\mu}}{G_{\mu}} 
- \frac{\delta M_{Z}^{2}}{M_{Z}^{2}}
+ \frac{\delta \alpha}{\alpha}
- \biggl( \frac{c_{\theta}^{2}-s_{\theta}^{2}}{c_{\theta}^{2}} \biggr)
\frac{\delta s_{\theta}^{2}}{s_{\theta}^{2}}.
\end{equation}
We note that $\Delta r_{Z}$ defined in Eq.(\ref{deltarz}) differs from 
the usual $\Delta \hat{r}_{Z}$ defined in the SM by an extra contribution 
due to the renormalization of $s_{\theta}^{2}$.

The counterterms for the Z-boson mass, $\delta M_{Z}^{2}$, and 
for the leptonic mixing angle, $\delta s_{\theta}^{2}$, 
are given by, respectively~\cite{Blank:1998qa},
\begin{eqnarray}
\delta{M_{Z}^{2}} & = & Re\biggl(\Pi^{ZZ}(M_{Z}^{2})\biggr)
\\
\frac{\delta s_{\theta}^{2}}{s_{\theta}^{2}} & = & 
Re \biggl[ \; \biggl(\frac{c_{\theta}}{s_{\theta}} \biggr)\; \biggr[
\frac{\Pi^{\gamma Z}(M_{Z}^{2})}{M_{Z}^{2}}
+ \frac{v_{e}^{2}-a_{e}^{2}}{a_{e}} \Sigma_{A}^{e}(m_{e}^{2}) 
\\
& & \qquad \qquad 
- \frac{v_{e}}{2s_{\theta}c_{\theta}}
\biggl( \frac{\Lambda_{V}^{Z\overline{e}{e}}(M_{Z}^{2})}{v_{e}}
-\frac{\Lambda_{A}^{Z\overline{e}{e}}(M_{Z}^{2})}{a_{e}} \biggr) \;
\biggr]\;
\biggr] . \nonumber
\end{eqnarray}
where $\Sigma_{A}^{e}$ is the axial part of the electron self-energy, 
$\Lambda_{V}^{Z\overline{e}{e}}$ and $\Lambda_{A}^{Z\overline{e}{e}}$ 
are the vector and axial-vector form factors of 
the vertex corrections to the $Z\overline{e}e$ coupling,  
$v_{e}$ is given by $v_{e}=1/2-2s_{\theta}^{2}$ 
and $a_{e}=1/2$. As the electron self-energy 
$\Sigma_{A}^{e}(m_{e}^{2})$ is suppressed by the 
small electron mass, it is negligible 
compared to other contributions. The vertex corrections 
$\Lambda_{V}^{Z\overline{e}{e}}(M_{Z}^{2})$ and 
$\Lambda_{A}^{Z\overline{e}{e}}(M_{Z}^{2})$
are both negligible as well, because both are proportional to the electron 
mass and thus are suppressed. In our analyses, we will therefore keep 
only contributions 
from $\Pi^{\gamma Z}(M_{Z}^{2})$ to 
$\delta s_{\theta}^{2}/s_{\theta}^{2}$. 

The electroweak radiative correction to the muon decay constant, 
$\delta G_{\mu}$, is due to the W-boson vacuum polarization, $\Pi^{WW}(0)$, 
and the vertex and box corrections, $\delta_{V-B}$. It is given by 
\begin{equation}
\frac{\delta G_{\mu}}{G_{\mu}} = -\frac{\Pi^{WW}(0)}{M_{W_{L}}^{2}} 
+ \delta_{V-B}.
\end{equation}
The vertex and box corrections, $\delta_{V-B}$, are small compared to the 
other correction~\cite{Blank:1998qa}, and is thus neglected in our analyses. 
The contribution due to the vacuum polarization of the photon, 
$\delta \alpha$, is given by
\begin{equation}
\frac{\delta \alpha}{\alpha} = \Pi^{\gamma \gamma \prime}(0) 
+ 2 (\frac{g_{V}^{e}-g_{A}^{e}}{Q_{e}}) \frac{\Pi^{\gamma Z}(0)}{M_{Z}^{2}}.
\end{equation}
Defining a short-hand notation $\Delta \hat{r}$, 
\begin{eqnarray}
\Delta \hat{r} & = & -\frac{\Delta s_{\theta}^{2}}{s_{\theta}^{2}}
+\frac{c^{2}s^{2}}{\sqrt{2}G_{\mu}f^{2}}
\\
& &
-\frac{Re(\Pi^{ZZ}(M_{Z}^{2}))}{M_{Z}^{2}} 
+ \Pi^{\gamma\gamma\prime}(0) + 2 (\frac{g_{V}-g_{A}}{Q_{e}})
\frac{\Pi^{\gamma Z}(0)}{M_{Z}^{2}}
\nonumber\\
&&
-\frac{c_{\theta}^{2}-s_{\theta}^{2}}{c_{\theta}s_{\theta}}
\frac{Re(\Pi^{\gamma Z}(M_{Z}^{2}))}{M_{Z}^{2}}, \nonumber
\end{eqnarray}
we can then write
\begin{equation}\label{eq:gmu}
s_{\theta}^{2}c_{\theta}^{2} = 
\frac{\pi \alpha(M_{Z})}{\sqrt{2}G_{\mu}M_{Z}^{2}\rho}
\biggl[ 1+\frac{\Pi^{WW}(0)}{M_{W_{L}}^{2}} + \Delta \hat{r} \biggr],
\end{equation}
Solving for $M_{W_{L}}^{2}$ and $\rho$ in Eq.(\ref{eq:rhodef}) and 
(\ref{eq:gmu}), we obtain a prediction for the physical W-boson mass
\begin{equation}\label{eq:mw}
M_{W_{L}}^{2} = \frac{1}{2} \biggl[ a (1+\Delta \hat{r}) + 
\sqrt{a^{2}(1+\Delta \hat{r})^{2} + 4a\Pi^{WW}(0)} \biggr]
\end{equation}
where $a \equiv \pi \alpha(M_{Z}^{2}) / \sqrt{2}G_{\mu} s_{\theta}^{2}$. 
The $\rho$ parameter is then predicted using Eq.(\ref{eq:rhodef}) with 
the $M_{W_{L}}^{2}$ value predicted in Eq.(\ref{eq:mw}).
Explicit expressions for the two point functions are given in the appendices.

We find that the one-loop contribution to $\Delta r_{Z}$ due to 
the SU(2) triplet scalar field, $\Phi$, scales as 
\begin{equation}
\Delta r_{Z}^{s}
\sim 
\frac{1}{16\pi^{2}} \frac{1}{v^{2}} (\frac{v^\prime}{v})^{2} M_{\Phi}^{2}
\quad .
\end{equation}
In the limit $v^{'} = 0$ while keeping $f$ fixed, which is equivalent to turning 
off the coupling $\lambda_{h \Phi h}$ in the Coleman-Weinberg potential, 
the one loop contribution due to the SU(2) triplet, 
$\Delta r_{Z}^{s}$, vanishes. 
The large $f$ limit of the scalar one-loop contribution, 
$\Delta r_{Z}^{s}$, vanishes depending upon how the limit 
$f \rightarrow \infty$ is taken.
As $f$ approaches infinity, the parameter $\mu^{2}$ (thus $v^{2}$) 
can be kept to be of the weak scale by fine-tuning the unknown coefficient 
in Eq.~\ref{eq:mu2} while all dimensionless parameters remain of order one. 
The scalar one-loop contribution in this limit does not de-couple because $M_{\Phi}^{2}$ 
increases as $f^{2}$ which compensates the $1/f^{2}$ suppression from 
$v^{\prime 2} /v^{2}$. In this case, the SM Higgs mass $m_{H}$ is of the weak 
scale $v$. 
On the other hand, without the fine-tuning mentioned above, 
$v$ can be held constant while varying $f$, 
if the quartic coupling $\lambda_{h^{4}}$ (thus $\lambda_{\Phi^{2}}$) 
approaches infinity as $f^{2}/v^{2}$. 
This can be done by taking $a \sim f^{2}/v^{2}$ while keeping 
$a^{\prime}$ finite and $s$ and $s^{\prime}$ having specific values. 
The scalar one-loop contribution then scales as
\begin{eqnarray}
\Delta r_{Z}^{s} 
& \sim & 
\frac{1}{v^{2}} (\frac{v^{\prime}}{v})^{2} M_{\Phi}^{2}
\\
& \sim & 
(\frac{1}{v^{2}}) 
(\frac{\lambda_{h\Phi h}}{\lambda_{\Phi^{2}}})^{2} \frac{v^{2}}{f^{2}} 
\lambda_{\Phi^{2}} f^{2}
\rightarrow \frac{\lambda_{h\Phi h}^{2}}{\lambda_{\Phi^{2}}} \quad .
\nonumber
\end{eqnarray}
Since the coupling constant $\lambda_{\Phi^{2}}$ must approach 
infinity in order to keep $v$ constant as we argue above, 
the scalar one-loop contribution 
$\Delta r_{Z}^{s}$ 
thus vanishes in the limit $f \rightarrow \infty$ with $v$ held fixed and
no fine tuning. 
In this case, $m_{H} \sim \mu$ scales with $f$.  
Of course, from the naturalness argument~\cite{Arkani-Hamed:2002qy} 
and unitarity constraint~\cite{Mahajan:2003vr},  
$f$ has an upper bound of a few TeV. The non-decoupling of heavy scalar 
fields has been noted before~\cite{Toussaint:1978zm,Senjanovic:1978ee}. 
A specific case of the de-coupling in the presence of the $SU(2)$ 
triplet Higgs in the LLH model is currently under 
investigation~\cite{chen:2003lh2}.

Blank and Hollik~\cite{Blank:1998qa} considered the complete one-loop radiative 
corrections to the electroweak observables in the Standard Model with an 
additional $SU(2)$ triplet with $Y=0$. 
They found large corrections to the $\rho$ parameter from one-loop corrections 
due to the triplet Higgs. Numerically, our results are consistent with theirs 
in appropriate limits.

\section{Numerical Results} 
 
We use the following experimentally measured values for the four 
input parameters~\cite{PDBook,lep03},
\begin{eqnarray}
G_{\mu} & = & 1.16639(1) \times 10^{-5} GeV^{-2}
\\ 
M_{Z} & = & 91.1876(21) \; GeV
\\ 
\alpha(M_{Z})^{-1} & = & 127.934 \pm 0.027
\\ 
s_{\theta}^{2} & = & 0.23150 \pm 0.00016. 
\end{eqnarray}
In addition, fermion masses and the Higgs boson mass are also unknown parameters. 
We use the following experimental values as inputs~\cite{PDBook,lep03}
\begin{equation}
m_{t} = 175 \, GeV
\end{equation}
and $m_{b}$ in $\overline{\mbox{MS}}$ scheme,
\begin{equation}
m_{b} =  3 \, GeV \, .
\end{equation}
And we choose
\begin{equation}
m_{H} =  120 \, GeV .
\end{equation}
In the Yukawa sector, there are 
two unknown parameters, the mixing angle $x_{L}$ between $t_{3}$ 
and $\tilde{t}$, and $\lambda$, which is defined as, 
\begin{equation}
\lambda \equiv \biggl({\lambda_{2}\over \lambda_{1}}\biggr)
\sqrt{\lambda_{1}^{2}+\lambda_{2}^{2}} \quad .
\end{equation}
We trade the top quark mass $m_{t}$ for $\lambda$, through the relation
\begin{equation}
m_{t} = \frac{\lambda x_{L}}{2^{1/4} G_{\mu}^{1/2}} \biggl[
1 + \frac{x_{L}}{2\sqrt{2}G_{\mu}f^{2}} \biggl( 1+x_{L} \biggr) 
\biggr] \; .
\end{equation}
and choose $m_{t}$ and $x_{L}$ as the two independent 
parameters in the Yukawa sector. 
In terms of the mass $m_{t}$ and the mixing angle $x_{L}$, the 
heavy top mass $M_{T}$ can be written as
\begin{equation}
M_{T} =  \frac{2^{1/4} G_{\mu}^{1/2} m_{t}}{\sqrt{x_{L}(1-x_{L})}} f 
\biggl[1 - \frac{x_{L}}{\sqrt{2}G_{\mu}f^{2}} 
\biggl( 1+x_{L} \biggr) \biggr] \quad .
\end{equation}

\begin{figure}
\psfrag{M(theory)-M(exp) (GeV)}[][]{\small $M_{\mbox{\tiny theory}}
-M_{\mbox{\tiny exp}}$ (GeV)}
\psfrag{x_L}[][]{\small $x_{L}$}
\psfrag{Mw  }[][]{\small $\delta M_{W_{L}}$(total) $\qquad $}
\psfrag{Mz  }[][]{\small $\delta M_{Z}$(total) $\qquad $}
\psfrag{Mwtree  }[][]{\small $\delta M_{W_{L}}$(tree) $\quad $}
\psfrag{deltaMw}[][]{\small 1 $\sigma$ limit on $\delta M_{W_{L}}$ (exp) $\qquad \qquad \qquad$ }
\psfrag{deltaMz}[][]{\small \small 1 $\sigma$ limit on $\delta M_{Z}$ (exp) $\quad \qquad \qquad \qquad $}
{\center
\includegraphics[scale=0.80,angle=270]{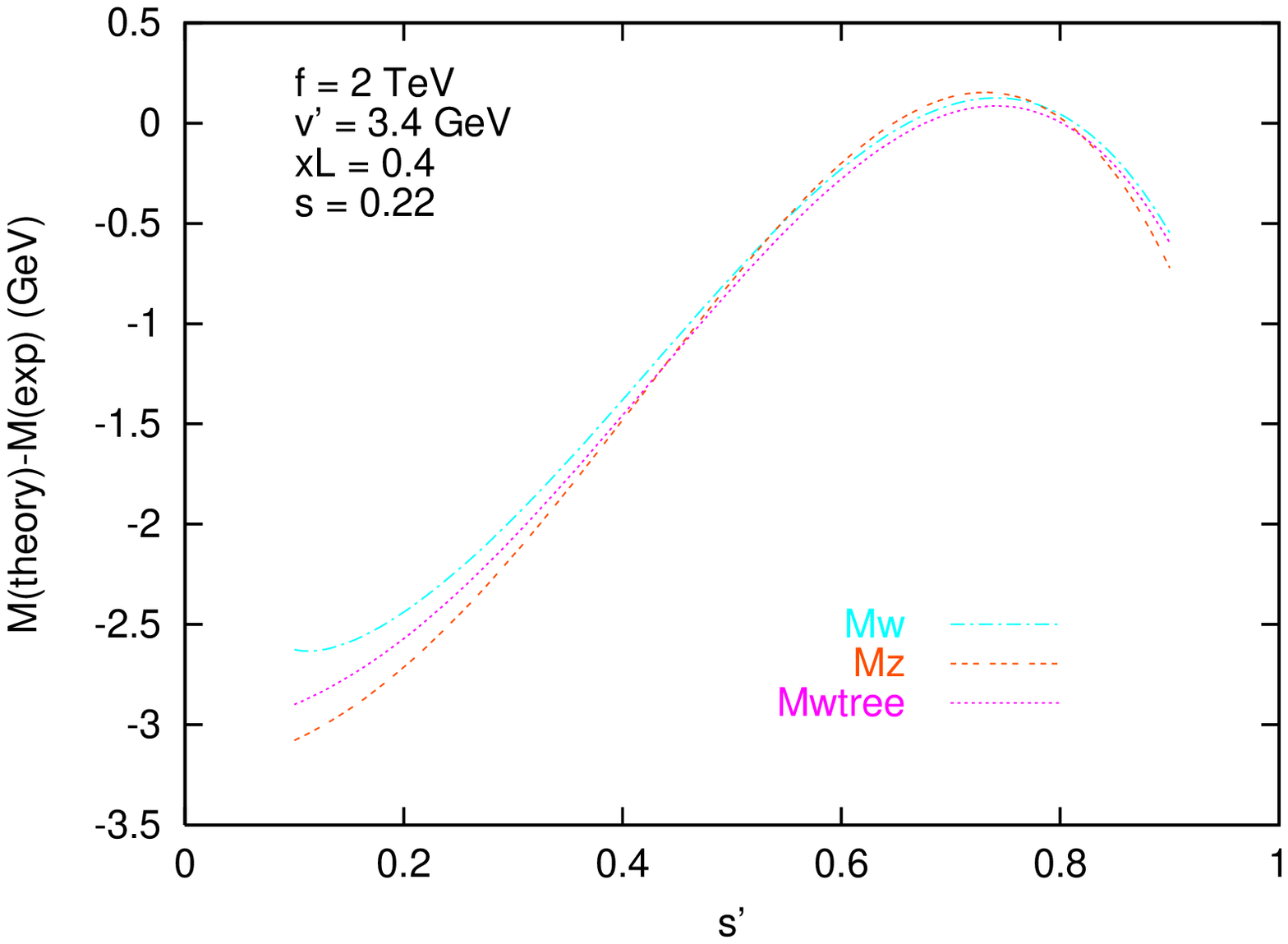}
\includegraphics[scale=0.80,angle=270]{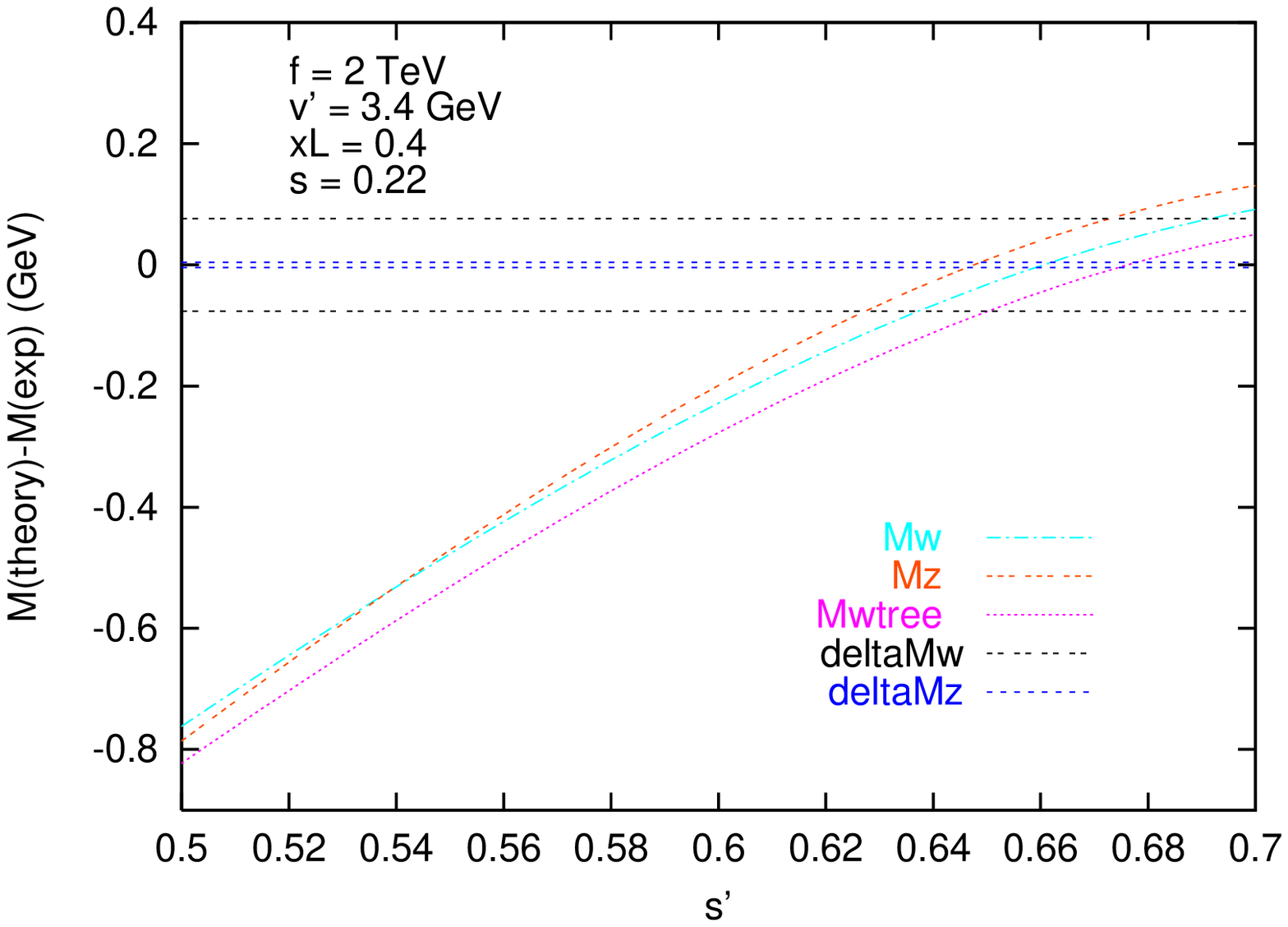}}
\caption{\label{fig:2000.1}
Prediction for $M_{W_{L}}$ as a function of the mixing angle $s^\prime$ at the tree level 
and the one-loop level. Also plotted is the correlation between $M_{Z}$ and 
$s^\prime$ for fixed $s$, $v^{\prime}$ and $f$. 
The cutoff scale $f$ in this plot is $2$ $TeV$, the $SU(2)$ triplet VEV 
$v^\prime = 3.4 \; GeV$, the mixing angle $s=0.22$, and $x_{L}=0.4$. 
}
\end{figure}

\begin{figure}
{\center
\includegraphics[scale=0.65]{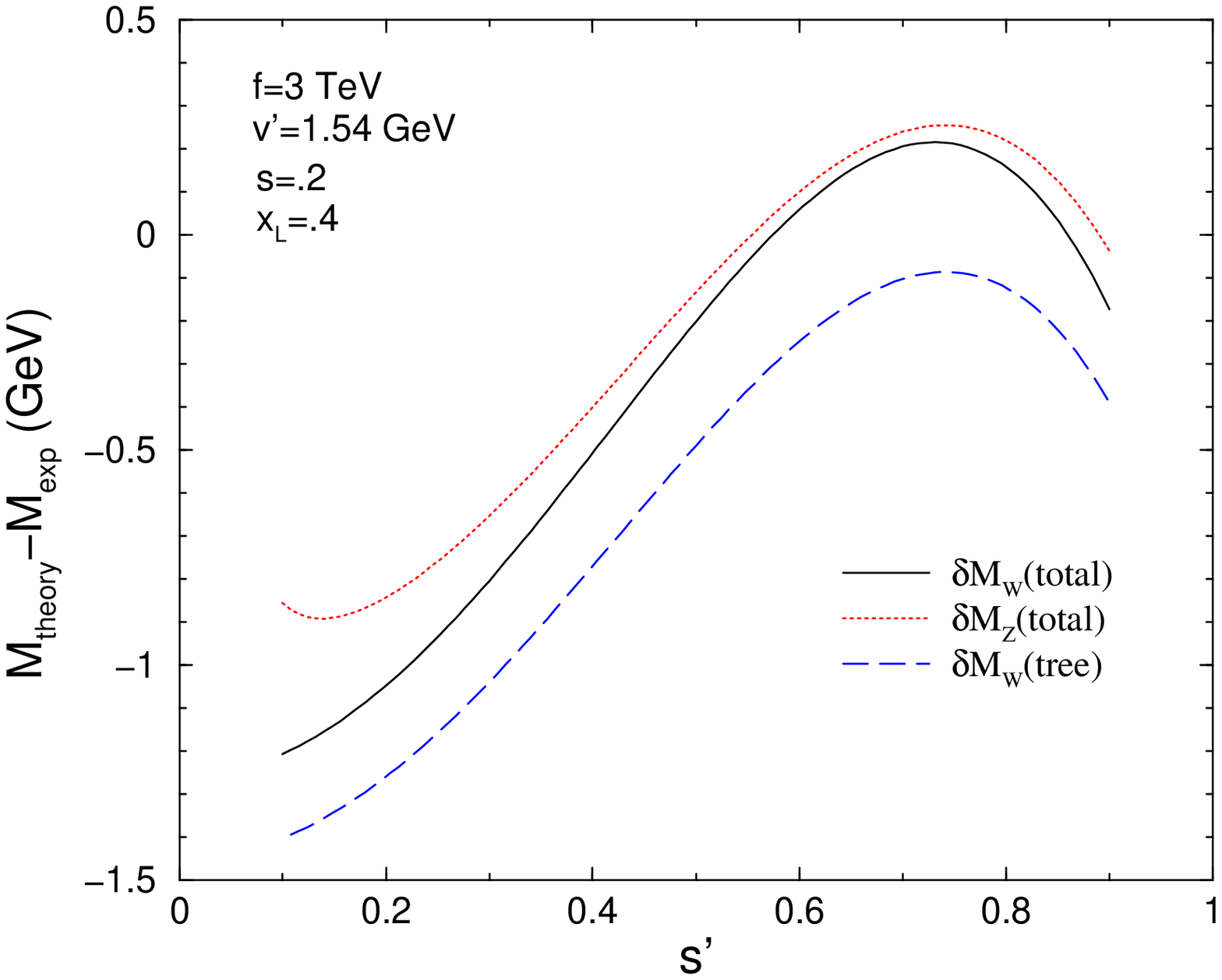}
\includegraphics[scale=0.65]{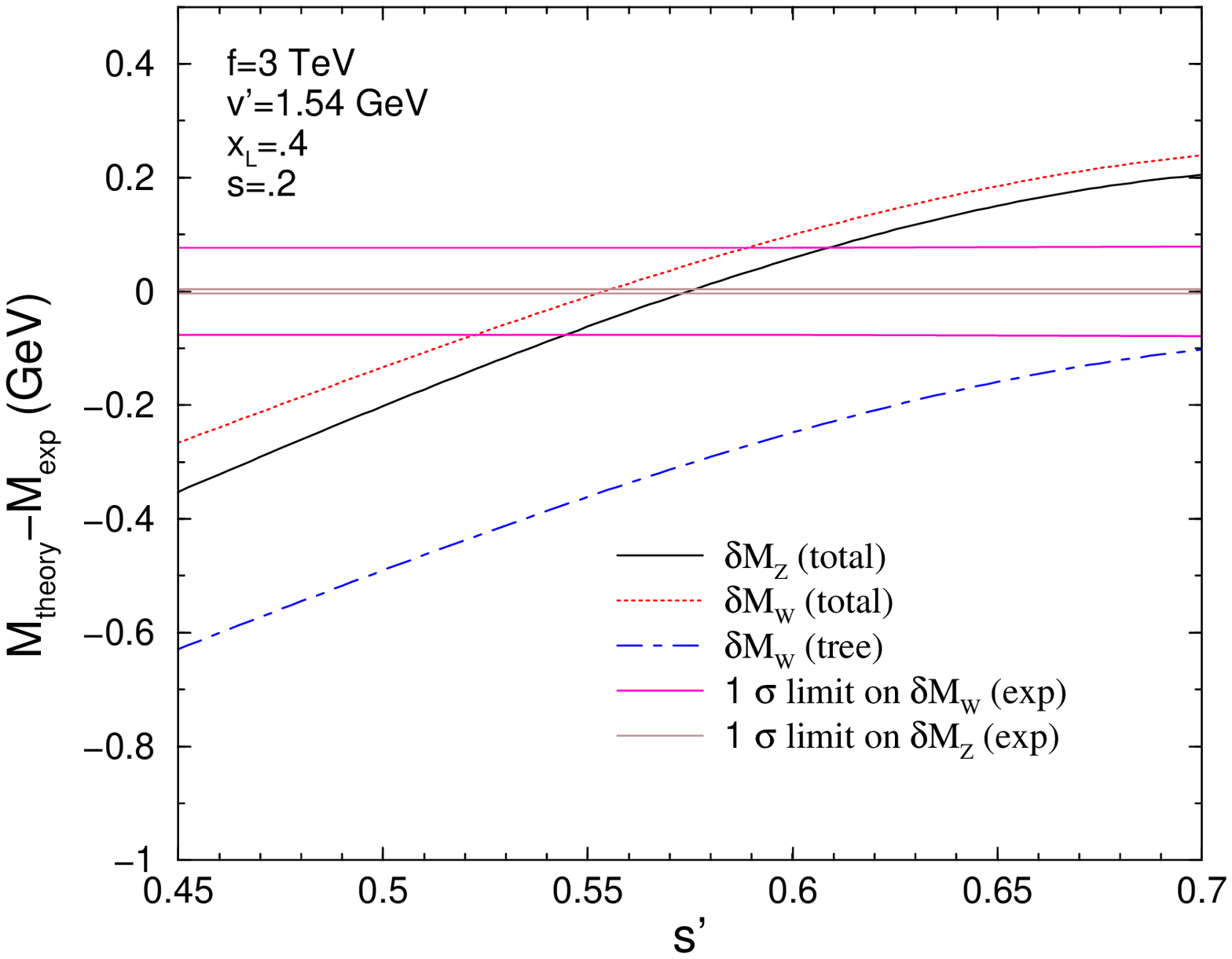}}
\caption{\label{fig:3000.1}
Prediction for $M_{W_{L}}$ as a function of the mixing angle $s^\prime$ at the tree level 
and the one-loop level. Also plotted is the correlation between $M_{Z}$ and 
$s^\prime$ for fixed $s$, $v^{\prime}$ and $f$. 
The cutoff scale $f$ in this plot is $3$ $TeV$, the $SU(2)$ triplet VEV 
$v^\prime = 1.54 \; GeV$, the mixing angle $s=0.2$, and $x_{L}=0.4$. 
}
\end{figure}

\begin{figure}
{\center
\includegraphics[scale=0.65]{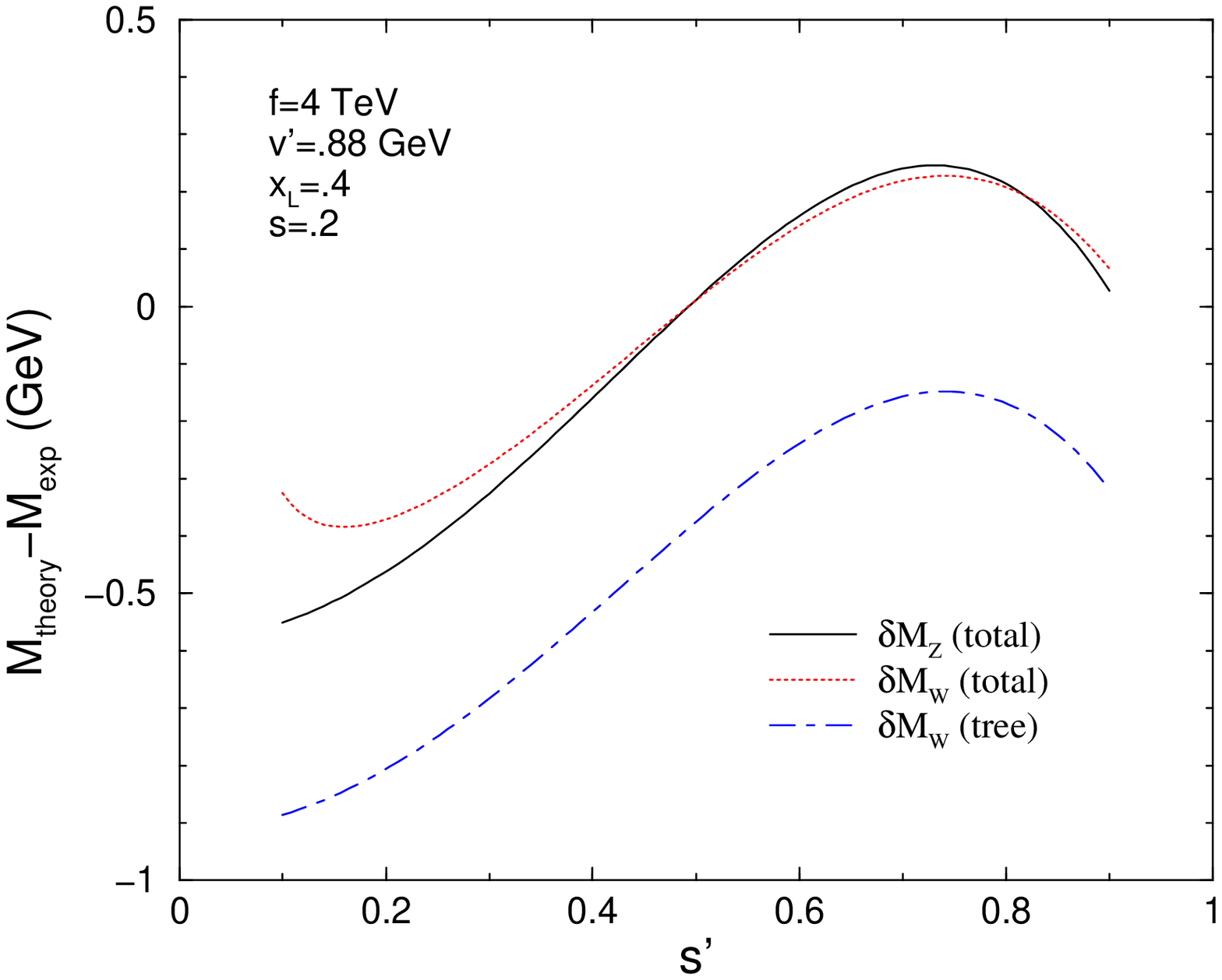}
\includegraphics[scale=0.65]{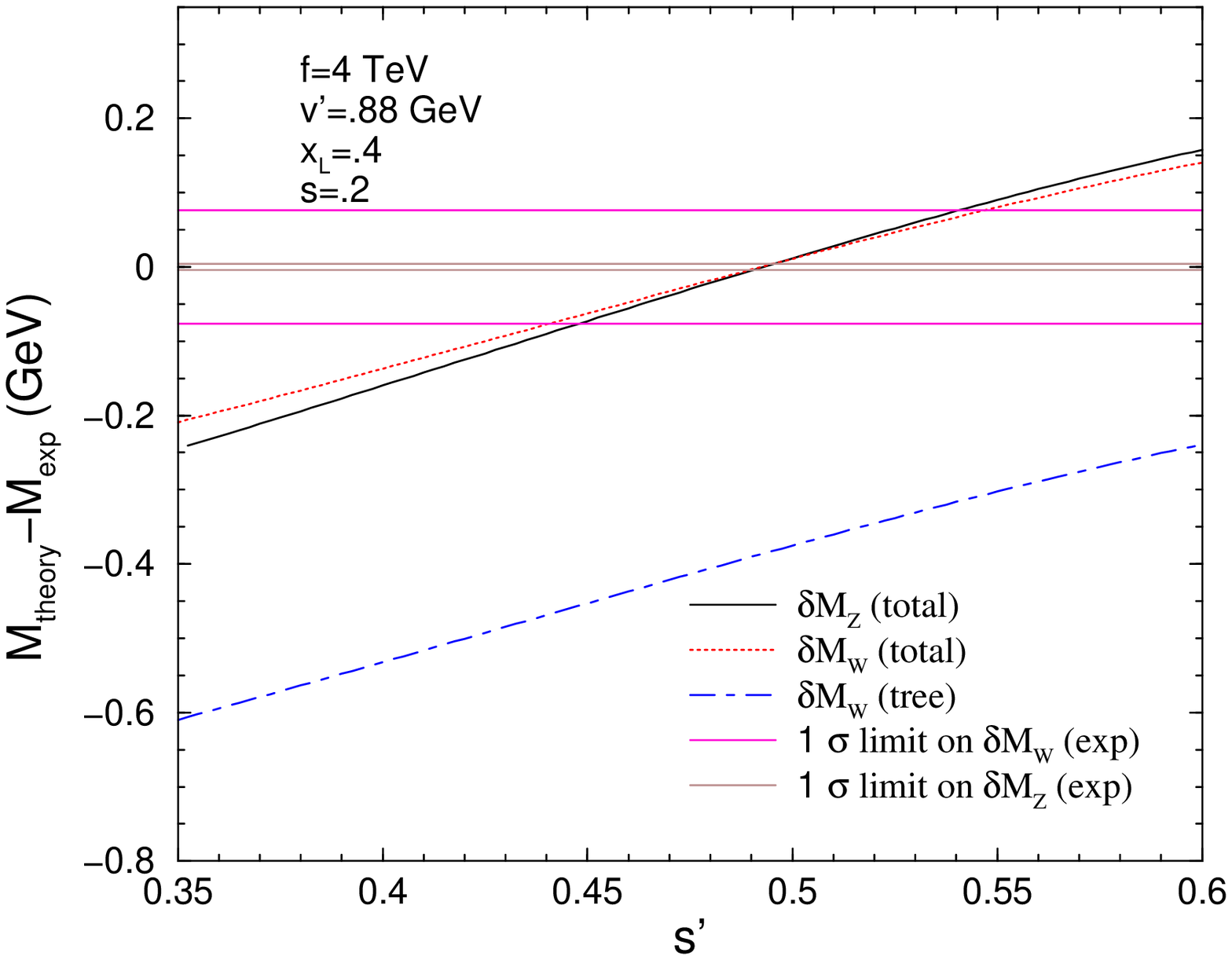}}
\caption{\label{fig:4000.1}
Prediction for $M_{W_{L}}$ as a function of the mixing angle $s^\prime$ at the tree level 
and the one-loop level. Also plotted is the correlation between $M_{Z}$ and 
$s^\prime$ for fixed $s$, $v^{\prime}$ and $f$. 
The cutoff scale $f$ in the plot is $4$ $TeV$, the $SU(2)$ triplet VEV 
$v^\prime = 0.88 \; GeV$, the mixing angle $s=0.2$, and $x_{L}=0.4$. 
}
\end{figure}

We analyze the dependence of the W-boson mass, $M_{W_{L}}$, on the mixing between 
$SU(2)_{1}$ and $SU(2)_{2}$, described by $s^{\prime}$, the mixing between $U(1)_{1}$ 
and $U(1)_{2}$, described by $s$, the mixing parameter in $t-T$ sector, $x_{L}$, 
and the VEV of the $SU(2)$, 
$v^{\prime}$. The predictions for $M_{W_{L}}$ with and without the one-loop contributions 
for $f=2$, $3$ and $4$ TeV are given in Figs.~\ref{fig:2000.1}, 
\ref{fig:3000.1} and \ref{fig:4000.1}, respectively.  These figures
demonstrate that a low value of $f$ ($f\sim 2~TeV$) is allowed by 
the experimental restrictions from the $W$ and $Z$ boson masses.  This
is because of the large effects of the one-loop corrections, in particular
the non-decoupling contributions of the scalar loops.  Figs.~\ref{fig:2000.1}, 
\ref{fig:3000.1} and \ref{fig:4000.1}, clearly demonstrate, however, that
in order to have experimentally acceptable gauge boson masses, 
the parameters of the
model must be quite finely tuned, regardless of the value of the scale $f$.

\begin{figure}
{\center
\includegraphics[scale=0.65]{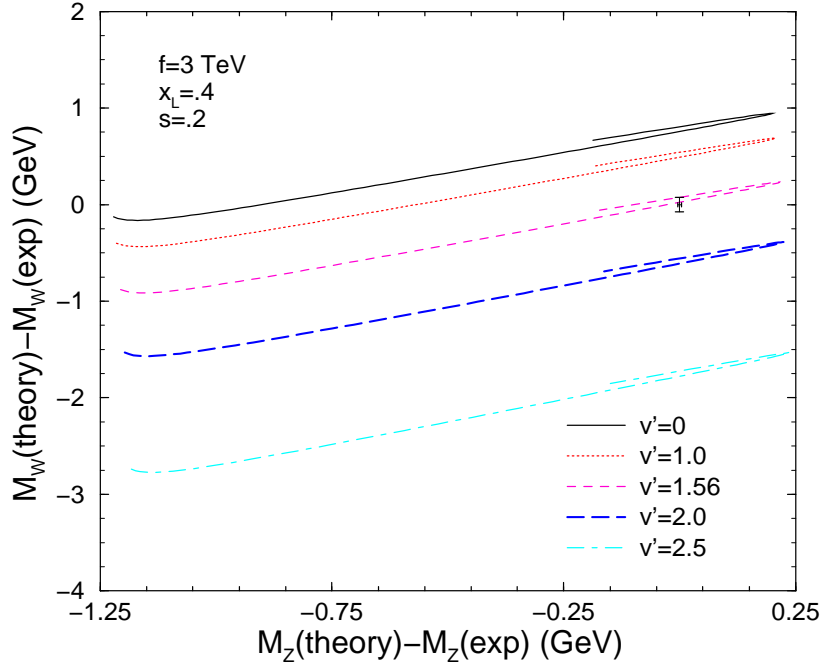}}
\caption{\label{fig:vpdep}Parametric plot of $M_{W_{L}}-M_{Z}$ in terms of $s^\prime$ for different 
values of the $SU(2)$ triplet VEV, $v^\prime = 0, \, 1.0, \, 1.56, \, 2.0$ and $2.5$. 
The cutoff scale $f$ is $3$ $TeV$, the mixing angle $s=0.2$, and $x_{L}=0.4$. The data 
point with error bars on $M_{W_{L}}$ and $M_{Z}$ is also shown. 
}
\end{figure}

\begin{figure}
\psfrag{x_L}[][]{\small $x_{L}$}
\psfrag{s}[][]{\small $s$}
\psfrag{f=2TeV}[][]{$\small f=2 \, TeV$}
\psfrag{f=3TeV}[][]{$\small f=3 \, TeV$}
\psfrag{f=4TeV}[][]{$\small f=4 \, TeV$}
\psfrag{f=5TeV}[][]{$\small f=5 \, TeV$}
\psfrag{f=6TeV}[][]{$\small f=6 \, TeV$}
{\center
\includegraphics[scale=0.32,angle=270]{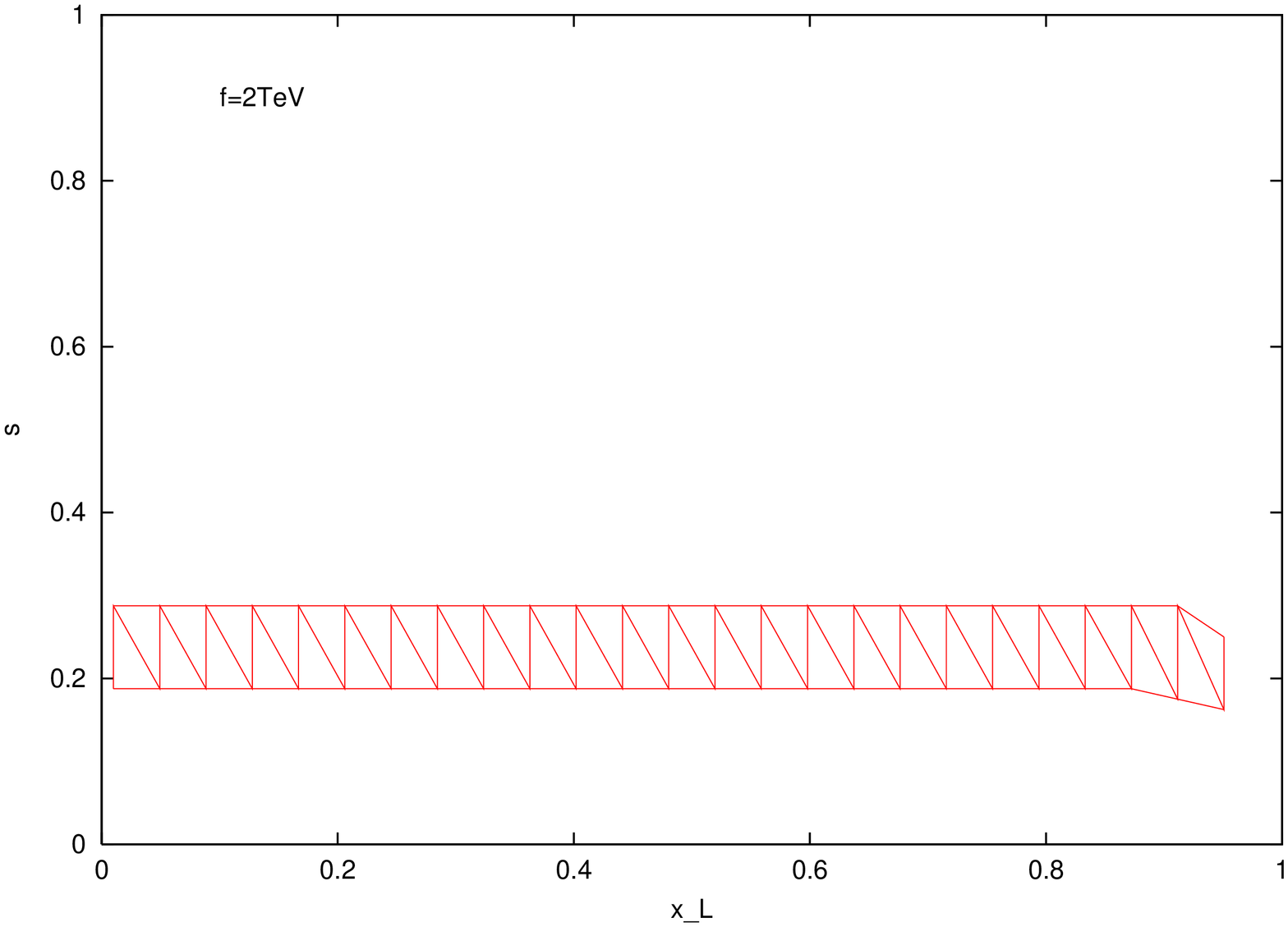}
\includegraphics[scale=0.32,angle=270]{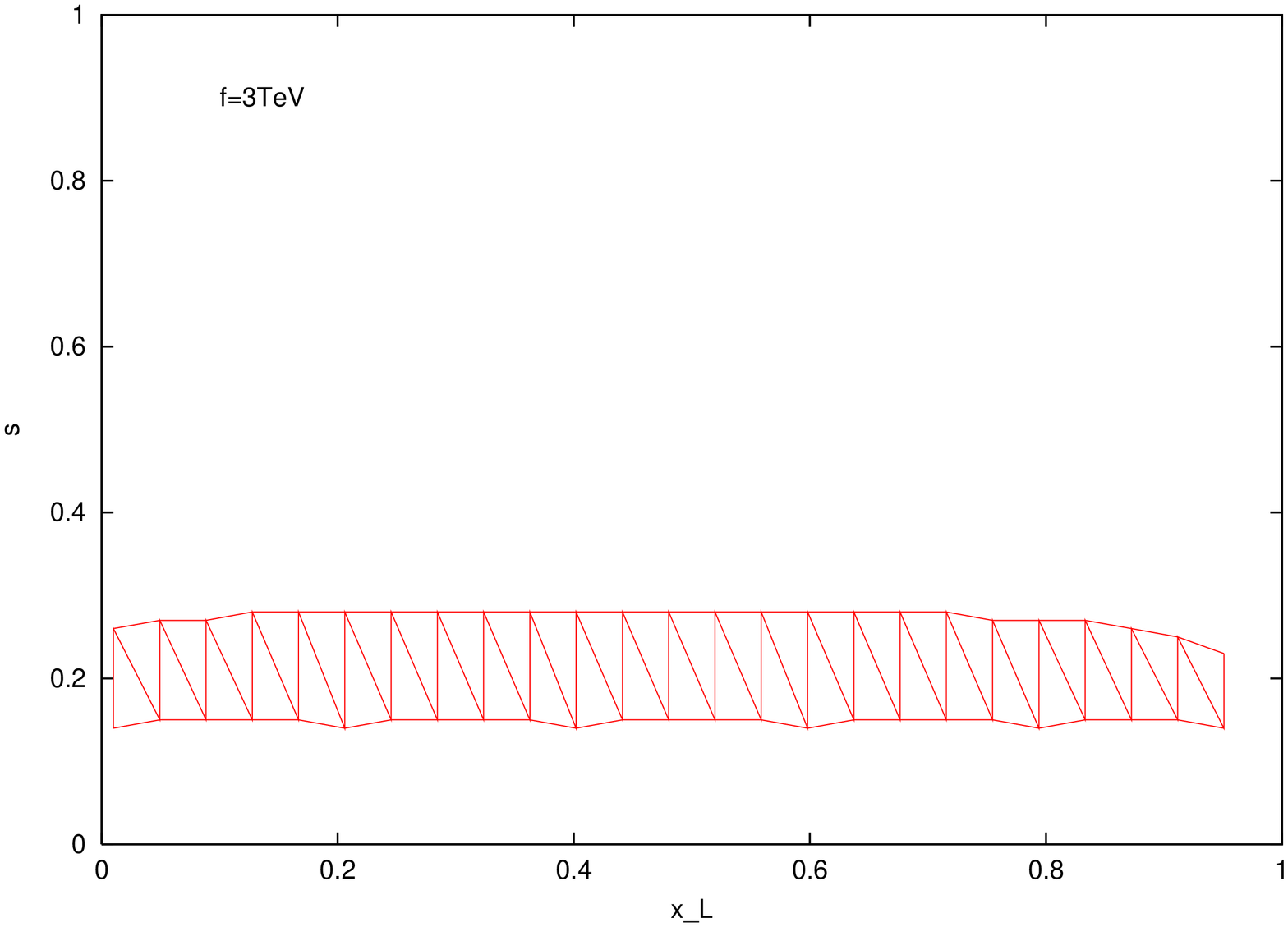}
\includegraphics[scale=0.32,angle=270]{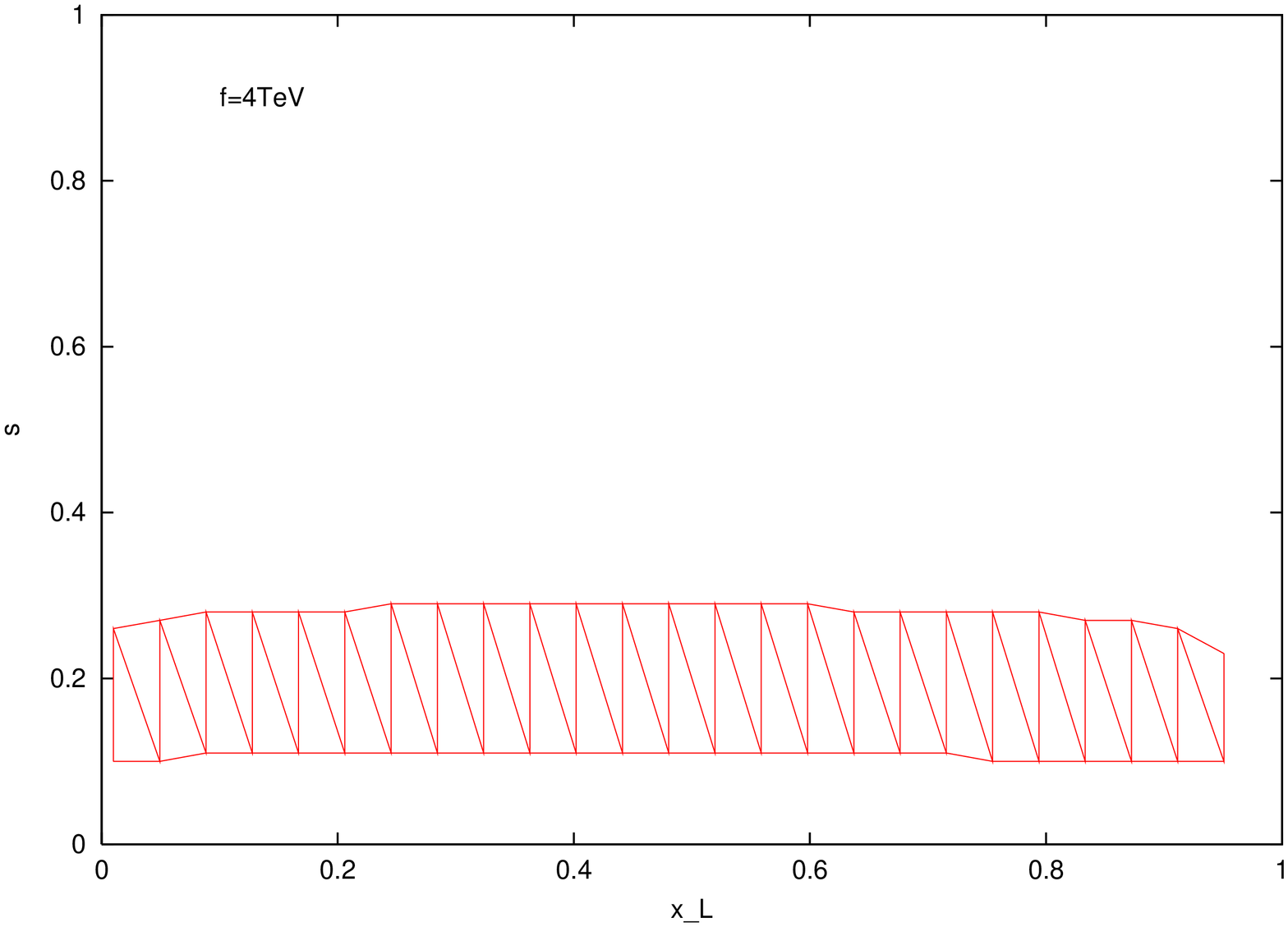}
\includegraphics[scale=0.32,angle=270]{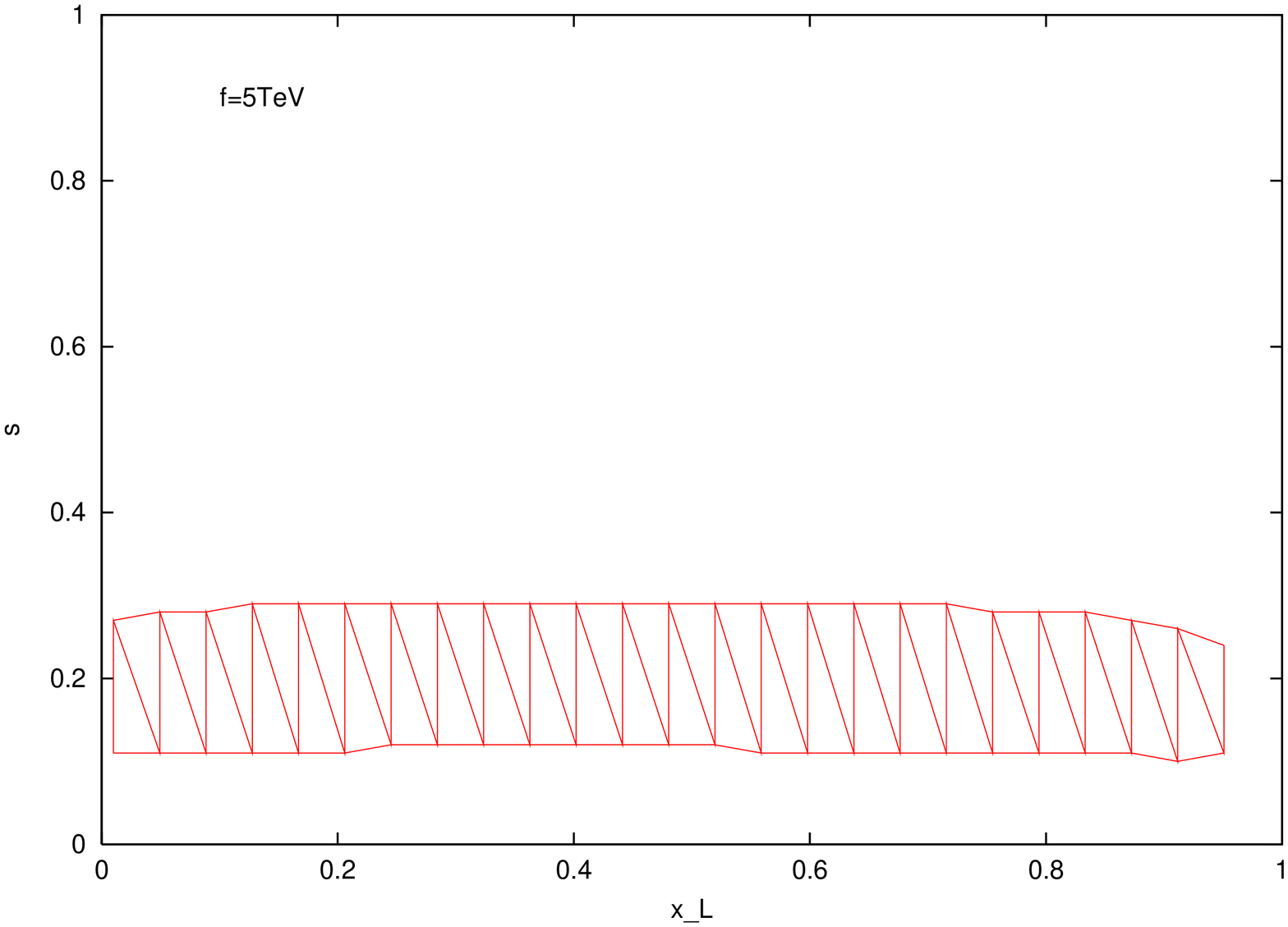}
\includegraphics[scale=0.32,angle=270]{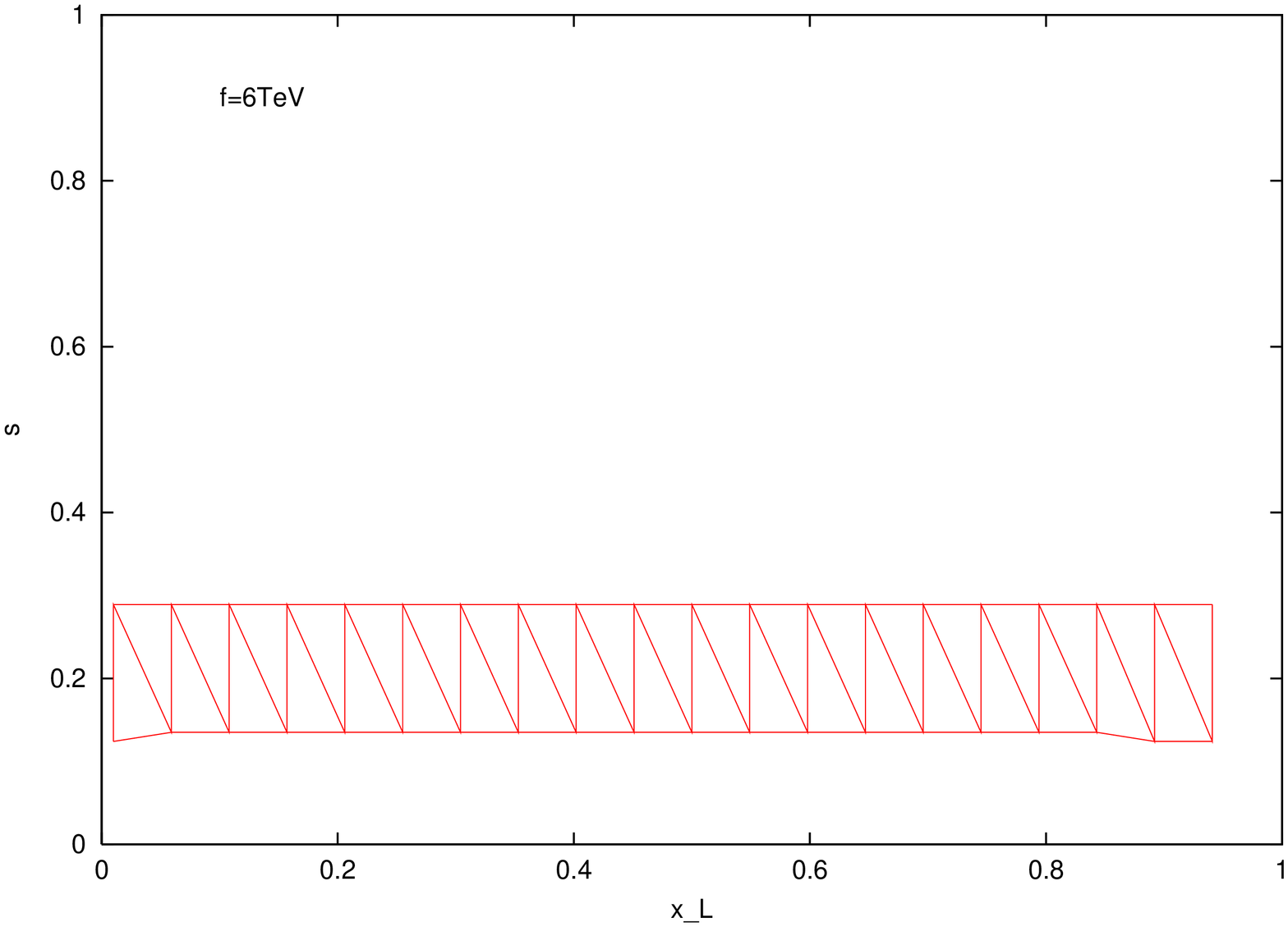}}
\caption{\label{fig1}
Allowed parameter space on the $(x_{L},s)$-plane, for $f=2, \; 3, \; 4, \; 5, \; 
6$ TeV. The triplet VEV $v^{\prime}$ is allowed to vary between $0$ and 
the upper bound 
given by Eq.~\ref{vvprelation}. For $f=(2,3,4,5,6)$ TeV, this bound is 
$v^{\prime}_{\mbox{\tiny max}} = (3.78, \, 2.52, \, 1.89, \, 1.51, \, 1.26)$ GeV.}
\end{figure}

\begin{figure}
\psfrag{v '}[][]{\small $v^{\prime}$ (GeV)}
\psfrag{s}[][]{\small $s$}
\psfrag{f=2TeV}[][]{$\small f=2 \, TeV$}
\psfrag{f=3TeV}[][]{$\small f=3 \, TeV$}
\psfrag{f=4TeV}[][]{$\small f=4 \, TeV$}
\psfrag{f=5TeV}[][]{$\small f=5 \, TeV$}
\psfrag{f=6TeV}[][]{$\small f=6 \, TeV$}
{\center
\includegraphics[scale=0.32,angle=270]{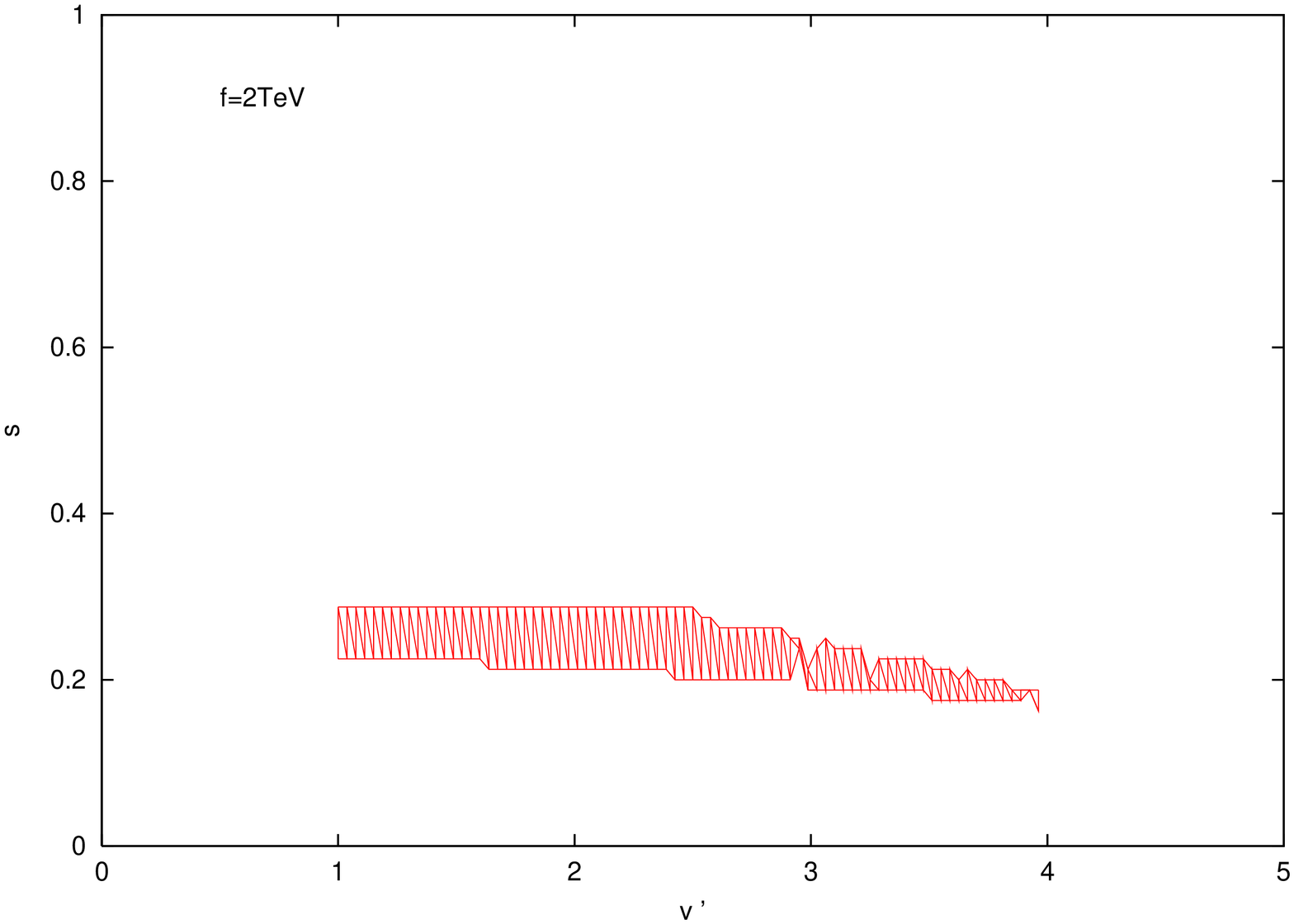}
\includegraphics[scale=0.32,angle=270]{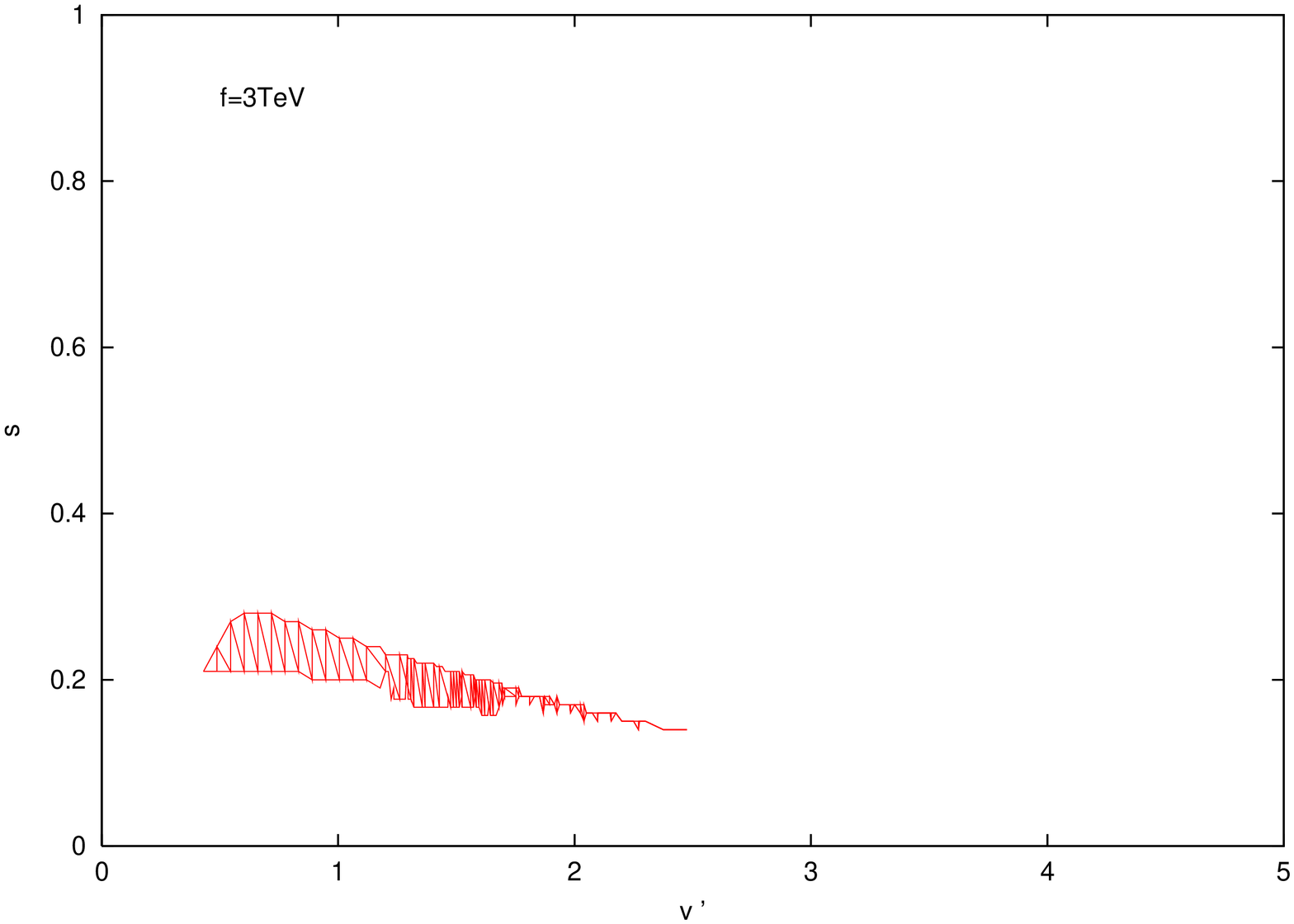}
\includegraphics[scale=0.32,angle=270]{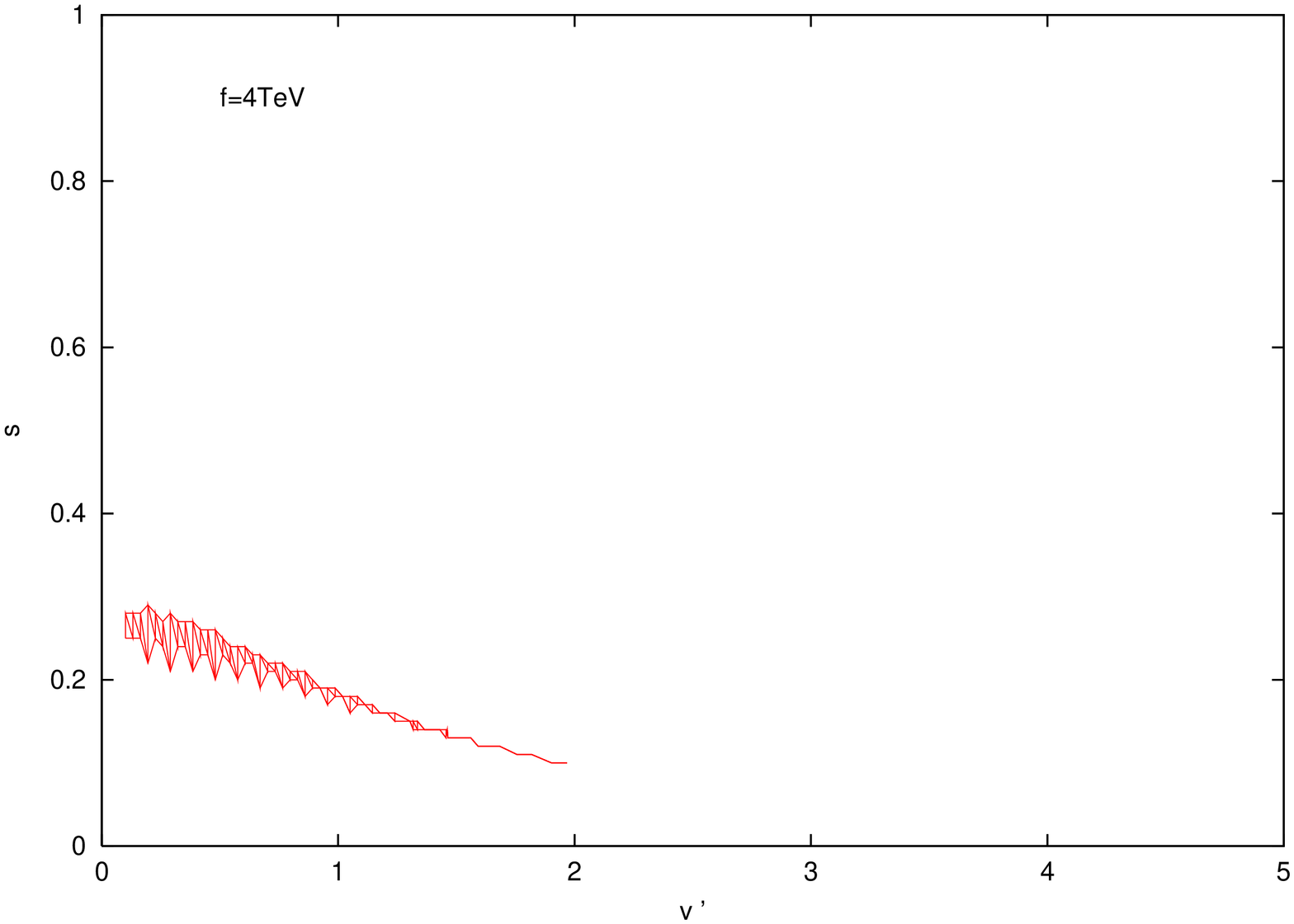}
\includegraphics[scale=0.32,angle=270]{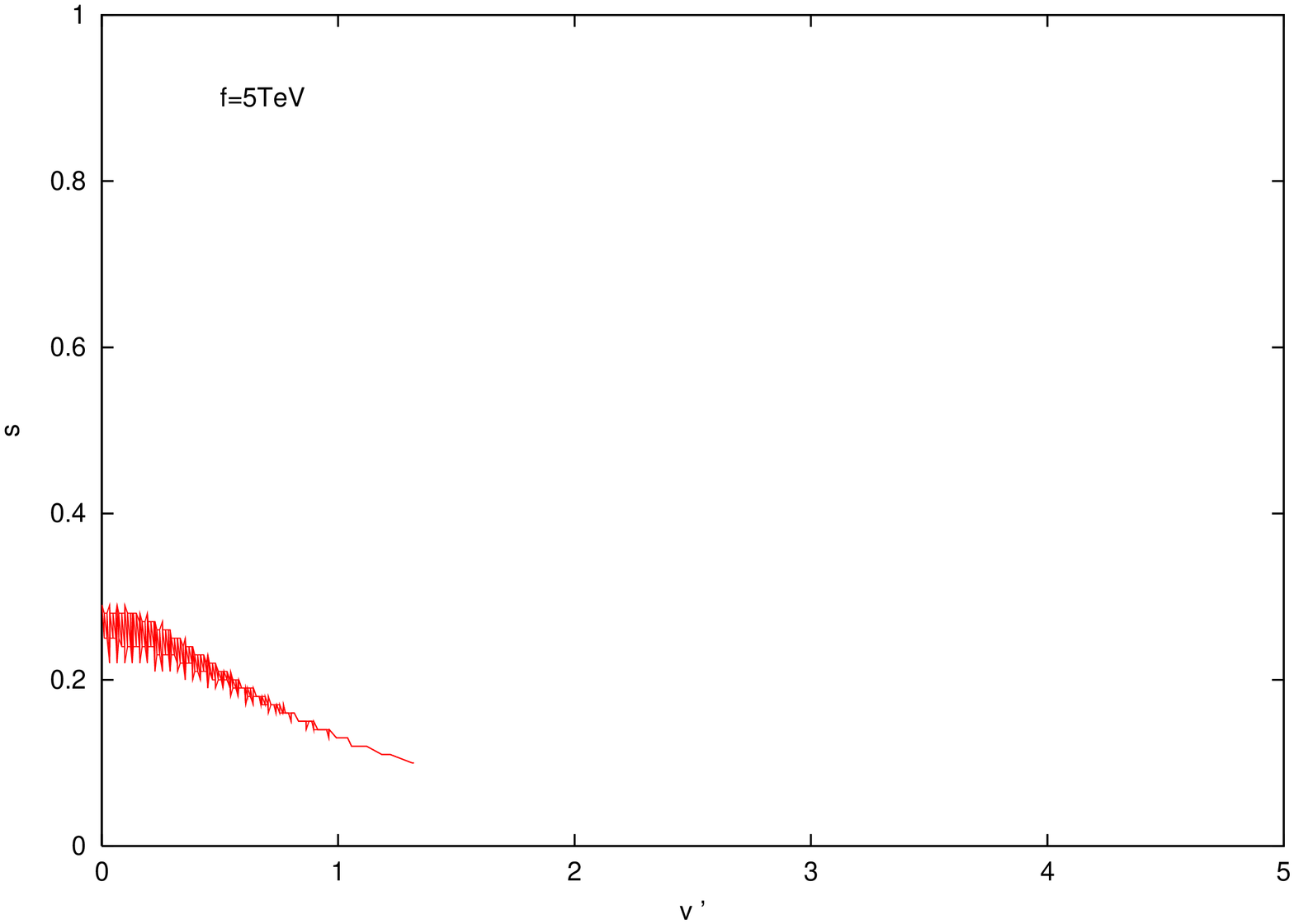}
\includegraphics[scale=0.32,angle=270]{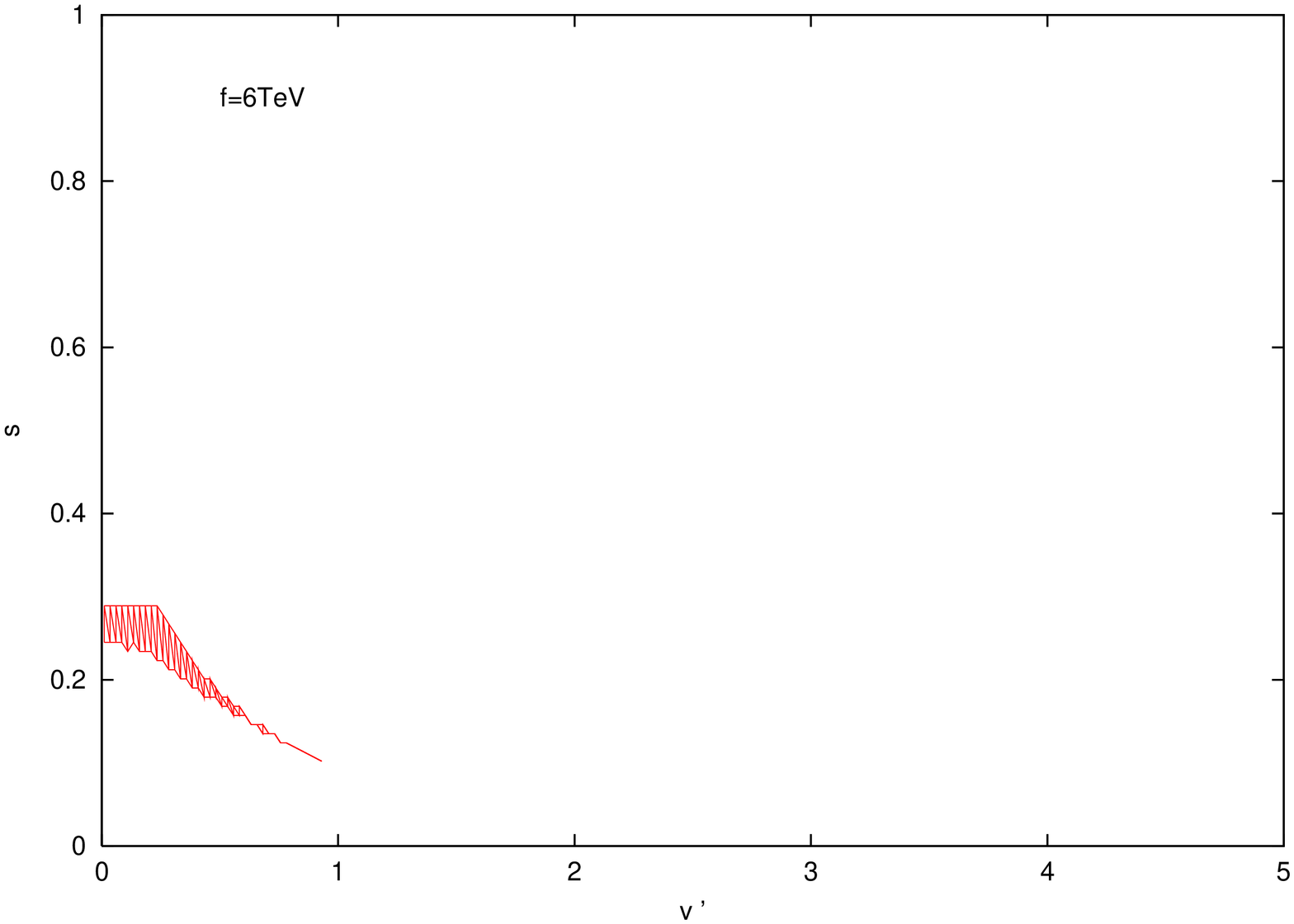}}
\caption{\label{fig2}
Allowed parameter space on the $(v^{\prime},s)$-plane for  
$f=2, \; 3, \; 4, \; 5, \; 6$ TeV. The mixing parameters $s^{\prime}$ and  
$x_{L}$ are allowed to vary between $0.01$ and $0.99$. For $f=(2,3,4,5,6)$ TeV, 
the upper bound given by Eq.~\ref{vvprelation} is 
$v^{\prime}_{\mbox{\tiny max}} = (3.78, \, 2.52, \, 1.89, \, 1.51, \, 1.26)$ GeV, respectively.
}
\end{figure}

\begin{figure}
\psfrag{s '}[][]{\small $s^{\prime} \;$}
\psfrag{v '}[][]{\small $v^{\prime} \qquad$}
{\center
\includegraphics[scale=0.8,angle=270]{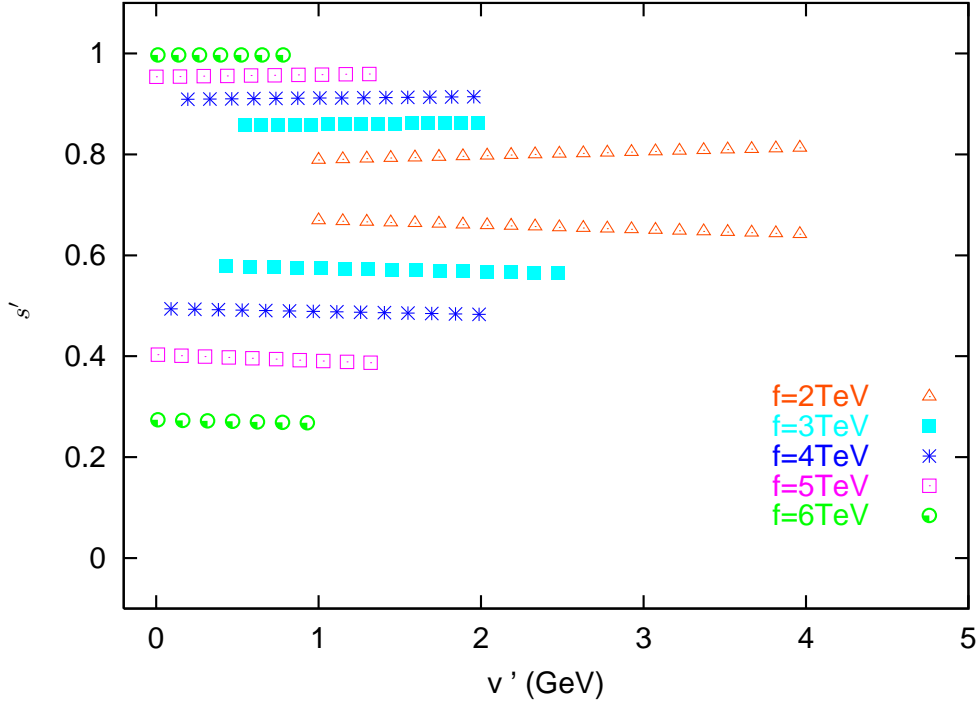}}
\caption{\label{fig3}
Allowed parameter space on the $(v^{\prime},s^{\prime})$-plane for  
$f=2, \; 3, \; 4, \; 5, \; 6$ TeV. The mixing parameters $s$ and $x_{L}$ 
are allowed to vary between $0.01$ and $0.99$.
}
\end{figure}

\begin{figure}
\psfrag{xL = 0.4}[][]{\small $x_{L} = 0.4$}
\psfrag{s ' = 0.5}[][]{\small $s^{\prime} = 0.5$}
\psfrag{s = 0.2}[][]{\small $s = 0.2$}
\psfrag{v ' = 1 GeV}[][]{\small $v^{\prime} = 1 GeV$}
\psfrag{Dtree  }[][]{$\scriptstyle \Delta_{\mbox{\tiny tree}}\qquad \qquad $  }
\psfrag{Df  }[][]{$\scriptstyle \Delta r_{Z}^{f} \qquad \qquad \qquad \; $}
\psfrag{Ds  }[][]{$\scriptstyle \Delta r_{Z}^{S} \qquad \qquad \qquad \; $}
\psfrag{D1loop  }[][]{$\scriptstyle \Delta \hat{r}_{Z} -\Delta_{\mbox{\tiny tree}} \qquad   $}
\psfrag{piww  }[][]{$\scriptstyle \Pi^{WW}(0)/M_{Z}^{2}  \qquad $}
{\center
\includegraphics[scale=0.8,angle=270]{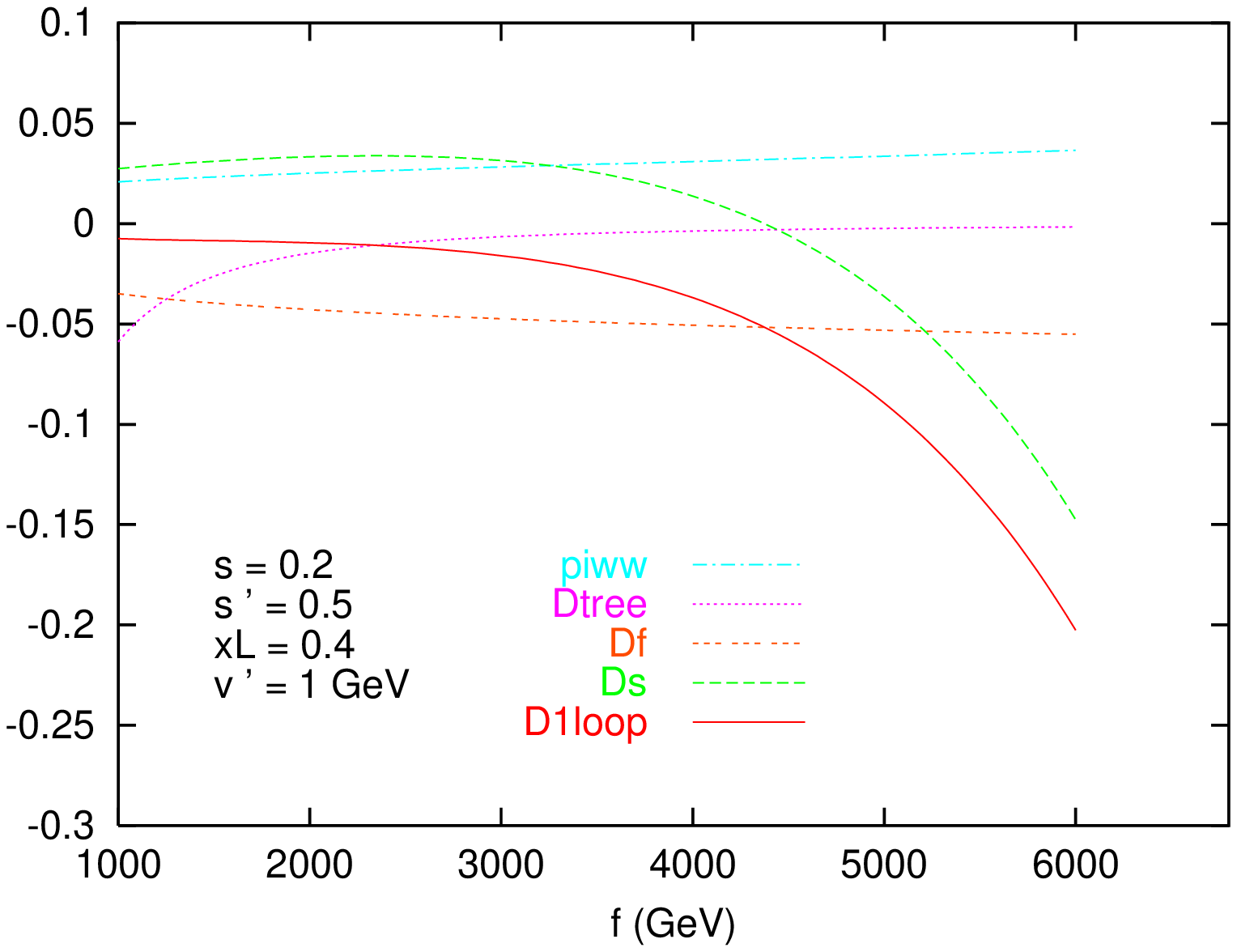}}
\caption{\label{fig4} The tree level correction, $\Delta_{\mbox{\tiny tree}}$, the 
fermionic and scalar contributions to the one loop correction, 
$\Delta r_{Z}^{f} $  
and $\Delta r_{Z}^{S}$,
the total one loop correction, $\Delta \hat{r}-\Delta_{\mbox{\tiny tree}}$, and $\Pi^{WW}(0)/M_{Z}^2$ 
as functions of the cutoff scale $f$ at fixed $s$, $s^{\prime}$, $x_{L}$ and $v^{\prime}$. }
\end{figure}

The importance of having a non-vanishing VEV, $v^{\prime}$, of the $SU(2)$ triplet is 
shown in Fig.~\ref{fig:vpdep}. 
The allowed parameter space on the $(x_{L},s)$-plane for various 
values of the cutoff scale is given in Fig.~\ref{fig1}. The allowed region on the 
$(v^{\prime},s)$-plane is given in Fig.~\ref{fig2}. In Fig.~\ref{fig3} the allowed 
region on the $(v^{\prime},s^{\prime})$-plane is shown. The non-decoupling of the 
$SU(2)$ triplet scalar field is shown in Fig.~\ref{fig4}.

Our analyses have shown that the model with low cutoff scale $f$ can still be in agreement 
with the experimental data, provided the VEV of the $SU(2)$ triplet scalar field 
is non-zero. This shows the importance of the $SU(2)$ triplet in placing the electroweak 
precision constraints. Constraint on the mixing parameter, $x_{L}$, is rather loose, as shown 
in Fig.~\ref{fig1}. The mixing parameter $s$ is bounded between $0.1$ and $0.3$; these bounds 
are insensitive to the cutoff scale, as shown in Fig.~\ref{fig1}. 

On the other hand, the prediction 
for $M_{W_{L}}$ is very sensitive to the values of $s^{\prime}$ as well as $v^{\prime}$.
The non-decoupling of the SU(2) triplet scalar field shown in Fig.~\ref{fig4} implies the 
importance of the inclusion of the scalar one-loop contributions in the analyses.
In the region below $f=4~TeV$, where the tree level corrections are large, 
 the vector boson self-energy is about half of the size of the tree level contributions, but 
with an opposite sign. (Other one-loop contributions roughly 
cancel among themselves in this region). Due to this cancellation 
between the tree level correction and the one-loop
correction, there is an allowed region of  parameter space
with low cutoff scale $f$.
Fig.~\ref{fig4} also shows that the tree level contribution of the LH
model get smaller as $f$ increases, as is expected.
In order to be consistent with experimental data, the triplet VEV $v^{\prime}$ 
must approach zero as $f$ goes to infinity, as shown in Fig.~\ref{fig2} and \ref{fig3}.  
The dependence on $m_{t}$ and $M_{T}$ is logarithmic as shown in Fig.~\ref{fig4}. 
This is consistent with the observation of Ref.~\cite{Czakon:1999ue,Czakon:1999ha}. 

\section{Conclusion}

In this paper we considered the logarithmically enhanced one-loop radiative 
corrections to the $\rho$ parameter, due to the additional heavy fermions 
and SU(2) triplet Higgs, including the contributions 
from both fermion and scalar loops.   We find the one-loop 
contributions, from both fermion and scalar sectors, can be 
comparable to the tree level correction to the $\rho$ parameter. 
In some cases, the one-loop contribution even dominates over 
the tree level correction due to the large
logarithmic enhancement of the loop corrections arising from
terms of ${\cal O}(\ln({f^2\over M_Z^2}))$. 

The fermion loop contribution dominates in the low cutoff 
scale region. On the other hand, the scalar loop contribution 
dominates in the high cutoff scale $f$ region; it grows with 
the cutoff scale $f$. This in turn implies an upper bound on 
the cutoff scale. The non-decoupling of the $SU(2)$ triplet 
is due to the fact that $M_{\Phi}^{2}$ scales as $f^{2}$ 
when the parameters in the theory are fine-tuned to fix $v$ 
at the weak scale and the
other parameters to be of order one.  Without this fine-tuning, the
triplet contributions do decouple for large $f$.  This non-decoupling
behavior of the scalar triplet will be further investigated in a future
publication~\cite{chen:2003lh2}.
Our results emphasize the need for a full one loop calculation.

\begin{acknowledgments}
We thank Sven Heinemeyer, K.T. Mahanthappa and Bill Marciano   
for useful correspondence and discussion, and Graham Kribs 
and Heather Logan for discussion and their very useful comments.
This work was supported by the U.S. Department of Energy 
under grant No. DE-AC02-76CH00016. 
\end{acknowledgments}

\appendix

\section{coupling constants in LLH Model}\label{sec:coupling}

We summarize in this section the relevant coupling constants for 
our calculation~\cite{Han:2003wu}. 
The gauge interaction of the fermions is given by
\begin{eqnarray}
\mathcal{L} & = & i\overline{\psi}_{1} \gamma_{\mu} (g_{V} + g_{A} \gamma_{5}) 
\psi_{2} X^{\mu}
\\
& = & i\overline{\psi}_{1} \gamma_{\mu} (c_{L} P_{L} + c_{R} P_{R}) 
\psi_{2} X^{\mu} \quad ,
\nonumber
\end{eqnarray}
where $P_{L} = \frac{1}{2}(1 - \gamma_{5})$ and 
$P_{R} = \frac{1}{2}(1 + \gamma_{5})$ are 
the usual projection operators.  
The gauge coupling constants of the fermions are given in Table \ref{ffx}.
\begin{table}[h]
\begin{tabular}{| l | l | l |}
\toprule
$X\overline{f}{f}$ & & 
\\
\hline
$W_{L} \overline{t} b$: \qquad & $c_{L} = \frac{g}{\sqrt{2}}$ 
& $c_{R} = 0$
\\
$W_{L} \overline{T} b$: & $c_{L} = -\frac{g}{\sqrt{2}} \frac{v}{f} x_{L}$ 
& $c_{R}=0$\\
\hline
$Z_{L} \overline{t} t$: & $g_{V} = \frac{g}{2c_{W}} (\frac{1}{2}-\frac{4}{3}s_{W}^{2})$ 
& $g_{A} = -\frac{g}{4c_{W}}$
\\
$Z_{L} \overline{b} b$: & $g_{V} = \frac{g}{2c_{W}}(-1/2+\frac{2}{3}s_{W}^{2}) $ 
& $g_{A} = \frac{g}{4c_{W}}$
\\
$Z_{L} \overline{T} T$: & $g_{V} = -\frac{2gs_{W}^{2}}{3c_{W}} $ 
& $g_{A} = \mathcal{O}(v^{2}/f^{2})$
\\
$Z_{L} \overline{T} t$: & $g_{V} =  \frac{gx_{L}}{4c_{W}}\frac{v}{f}$ 
& $g_{A} = -\frac{gx_{L}}{4c_{W}} \frac{v}{f}$
\\
\hline
$A_{L} \overline{f}{f}$: & $g_{V} = eQ_{f}$ & $g_{A} = 0$
\\
\botrule
\end{tabular}
\caption{\label{ffx} Relevant coupling constants 
$X\overline{f}f$. As $M_{T}$ is of order $f$, 
gauge coupling constants for $T$ and $\overline{T}$ 
must be expanded to order $v/f$. 
For the coupling constants of the light fermions, 
we retain only the order $(v/f)^{0}$ 
terms. $Q_{f}$ is the electric charge of fermion $f$: 
$Q_{t} = Q_{T} = +2/3$, $Q_{b} = -1/3$. 
We make the approximation that $V_{tb}^{\mbox{\tiny SM}}
=1$~\cite{Han:2003wu}.}
\end{table}

The gauge coupling constants of the scalar fields are given in 
Table \ref{XXSS}, \ref{XSS} 
and \ref{XXS}. The parameters $s_{0}$, $s_{p}$ and $s_{+}$ describe the 
mixing in the 
neutral CP-even scalar, pseudoscalar and singly charged sectors, respectively.  
To leading order in $v^\prime \over v$ they are given by~\cite{Han:2003wu},
\begin{eqnarray}
s_{0} & \simeq & 2\sqrt{2} \frac{v^{'}}{v}
\label{eq:s1}
\\
s_{p} & = & \frac{2\sqrt{2}v^{'}}{\sqrt{v^{2}+8v^{'2}}} \simeq 2\sqrt{2} \frac{v^{'}}{v}
\label{eq:s2}\\
s_{+} & = & \frac{2v^{'}}{\sqrt{v^{2}+4v^{'2}}} \simeq 2 \frac{v^{'}}{v} \quad .
\label{eq:s3}
\end{eqnarray}

\begin{table}
\begin{tabular}{| l | l || l | l |}
\toprule
$XXSS$ & $C_{\scriptstyle XXSS}$ &
$XXSS$ & $C_{\scriptstyle XXSS}$
\\
\hline
$W_{L}^{+} W_{L}^{-} H H$ & $\frac{g^{2}}{2}$ &
$Z_{L} Z_{L} HH$ & $\frac{g^{2}}{2c_{W}^{2}}$
\\
$W_{L}^{+} W_{L}^{-} \Phi^{0} \Phi^{0}$ & $g^{2}$ &
$Z_{L} Z_{L} \Phi^{0} \Phi^{0}$ & $2 \frac{g^{2}}{c_{W}^{2}}$
\\
$W_{L}^{+} W_{L}^{-} \Phi^{P} \Phi^{P}$ & $g^{2}$ &
$Z_{L} Z_{L} \Phi^{P} \Phi^{P}$ & $2 \frac{g^{2}}{c_{W}^{2}}$
\\
$W_{L}^{+} W_{L}^{-} \Phi^{+} \Phi^{-}$ & $2g^{2}$ &
$Z_{L} Z_{L} \Phi^{+} \Phi^{-}$ & $2 \frac{g^{2}}{c_{W}^{2}} s_{W}^{4}$
\\
$W_{L}^{+} W_{L}^{-} \Phi^{++} \Phi^{--}$ & $g^{2}$ &
$Z_{L} Z_{L} \Phi^{++} \Phi^{--}$ & $2 \frac{g^{2}}{c_{W}^{2}} (1-2s_{W}^{2})^{2}$
\\
$A_{L}A_{L}\Phi^{+}\Phi^{-}$ & $2e^{2}$ &
$A_{L} A_{L} \Phi^{++}\Phi^{--}$ & $8e^{2}$
\\
$A_{L}Z_{L}\Phi^{+}\Phi^{-}$ & $-2e \frac{g}{c_{W}} s_{W}^{2}$ &
$A_{L}Z_{L}\Phi^{++}\Phi^{--}$ & $4e \frac{g}{c_{W}} (1-2s_{W}^{2})$
\\
\botrule
\end{tabular}
\caption{\label{XXSS} Relevant gauge coupling constants of the scalar fields, 
$C_{\scriptstyle XXSS}$~\cite{Han:2003wu}.}
\end{table}

\begin{table}
\begin{tabular}{| l | l || l | l || l | l |}
\toprule
$ XSS$ & $C_{\scriptstyle XSS}$ & 
$ XSS$ & $C_{\scriptstyle XSS}$ &
$ XSS$ & $C_{\scriptstyle XSS}$ \\
\hline
$W_{L}^{+} H \Phi^{-}$ & $-\frac{g}{2}(\sqrt{2}s_{0}-s_{+})$ &
$Z_{L} H \Phi^{P}$ & $\frac{g}{2c_{W}} (s_{p}-2s_{0})$ &
$A_{L} \Phi^{+} \Phi^{-}$ & $-e$
\\
$W_{L}^{+} \Phi^{0} \Phi^{-}$ & $-\frac{g}{\sqrt{2}}$ &
$Z_{L} \Phi^{0} \Phi^{P}$ & $-\frac{g}{c_{W}}$ &
$A_{L} \Phi^{++} \Phi^{--}$ & $-2e$
\\
$W_{L}^{+} \Phi^{P} \Phi^{-}$ & $\frac{g}{\sqrt{2}}$ &
$Z_{L} \Phi^{+} \Phi^{-}$ & $\frac{g}{c_{W}}s_{W}^{2}$
&&
\\
$W_{L}^{+} \Phi^{+} \Phi^{--}$ & $-g$ &
$Z_{L} \Phi^{++} \Phi^{--}$ & $-\frac{g}{c_{W}}(1-2s_{W}^{2})$
&&
\\
\botrule
\end{tabular}
\caption{\label{XSS} Relevant gauge coupling constants of the scalar fields, 
$C_{\scriptstyle XSS}$~\cite{Han:2003wu}.}
\end{table}

\begin{table}
\begin{tabular}{| l | l || l | l |}
\toprule
$X_{1}X_{2}S$ & $C_{X_{1}}C_{X_{2}}S$ &
$X_{1}X_{2}S$ & $C_{X_{1}}C_{X_{2}}S$
\\
\hline
$W_{L}^{+} W_{L}^{-} H$ & $\frac{1}{2} g^{2} v$ &
$Z_{L} Z_{L} H$ & $\frac{1}{2} \frac{g^{2}}{c_{W}^{2}} v$
\\
$W_{L}^{+} W_{H}^{-} H$ & $-\frac{1}{2} g^{2} \frac{c^{2}-s^{2}}{2sc} v$ &
$Z_{L} A_{H} H$ & $-\frac{1}{2} \frac{gg^{'}}{c_{W}} \frac{c^{'2}-s^{'2}}{2s^{'}c^{'}} v$
\\
$W_{L}^{+} W_{L}^{-} \Phi^{0}$ & $-\frac{1}{2} g^{2} (s_{0} v - 2\sqrt{2}v^{'}) $ &
$Z_{L} Z_{L} \Phi^{0}$ & $-\frac{1}{2} \frac{g^{2}}{c_{W}^{2}} 
(s_{0} v - 4\sqrt{2}v^{'}) $ 
\\
$W_{L}^{+} W_{H}^{-} \Phi^{0}$ & $\frac{g^2}{4} \frac{c^2-s^2}{s c} 
(s_0v-2\sqrt{2}v^\prime)$
&
$Z_{L} Z_{H} \Phi^{0}$ & $\frac{1}{2} \frac{g^{2}}{c_{W}} 
\frac{c^{2} - s^{2}}{2sc} (s_{0} v - 4\sqrt{2}v^{'}) $
\\
$W_{L}^{+} A_{L} \Phi^{-}$ & 0 &
$Z_{L}Z_{H}H$ & 
$-\frac{1}{2} \frac{g^{2}}{c_{W}}\frac{c^{2} - s^{2}}{2sc} v$
\\
$W_{L}^{+} A_{H} \Phi^{-}$ & $-\frac{1}{2} gg^{'} \frac{c^{'2}-s^{'2}}{2s^{'}c^{'}}
(s_{+} v - 4 v^{'})$ &
$Z_{L}A_{H}\Phi^{0}$ &
$\frac{1}{2}\frac{gg^{'}}{c_{W}}\frac{c^{'2}-s^{'2}}{2s^{'}c^{'}}
(s_{0} v - 4\sqrt{2}v^{'})$
\\
$W_{L}^{+} W_{L}^{+} \Phi^{--}$ &
$2g^{2}v^{'}$ &
$W_{L}^{+} Z_{L} \Phi^{-}$ & $ -\frac{g^{2}}{c_{W}} v^{'}$ 
\\
$W_{L}^{+} W_{H}^{+} \Phi^{--}$ &
$-2g^{2} \frac{c^{2} - s^{2}}{2sc} v^{'}$ &
$W_{L}^{+} Z_{H} \Phi^{-}$ &
$g^{2} \frac{c^{2} - s^{2}}{2sc} v^{'}$
\\
$W_{H}^{+} Z_{L}^{+} \Phi^{-}$ &
$\frac{g^{2}}{c_w}  \frac{c^{2} - s^{2}}{2sc} v^{'}$ & &
\\
\botrule
\end{tabular}
\caption{\label{XXS} Relevant gauge coupling constants of the scalar fields, 
$C_{\scriptstyle XXS}$~\cite{Han:2003wu}.}
\end{table}

\section{One-Loop Integrals}

The one-loop integrals are decomposed in terms of 
Passarino-Veltman~\cite{Passarino:1979jh}
functions which are defined in $n=4-2\epsilon$ dimensions,
\begin{eqnarray}
Q^{4-n} \int \frac{d^{n} k}{(2\pi)^{n}} \frac{1}{k^{2}-m^{2}+i\epsilon}
& \equiv & 
\frac{i}{16\pi^{2}} A_{0}(m^{2})  
\\
Q^{4-n} \int \frac{d^{n} k}{(2\pi)^{n}} 
\frac{1}{(k^{2}-m_{1}^{2}+i\epsilon)((k-p)^{2}-m_{2}^{2}+i\epsilon)}
& \equiv & 
\frac{i}{16\pi^{2}} B_{0}(p^{2},m_{1}^{2},m_{2}^{2})  
\\
Q^{4-n} \int \frac{d^{n} k}{(2\pi)^{n}} 
\frac{k_{\mu}}{(k^{2}-m_{1}^{2}+i\epsilon)((k-p)^{2}-m_{2}^{2}+i\epsilon)}
& \equiv &  
\frac{i}{16\pi^{2}} p_{\mu} B_{1}(p^{2},m_{1}^{2},m_{2}^{2}) 
\\
Q^{4-n} \int \frac{d^{n} k}{(2\pi)^{n}} 
\frac{k_{\mu}k_{\nu}}{(k^{2}-m_{1}^{2}+i\epsilon)((k-p)^{2}-m_{2}^{2}+i\epsilon)}
& \equiv & \\ 
\qquad \frac{i}{16\pi^{2}} [ g_{\mu\nu} B_{22} (p^{2},m_{1}^{2},m_{2}^{2}) & + & 
p_{\mu}p_{\nu}B_{11}(p^{2},m_{1}^{2},m_{2}^{2})], \nonumber  
\end{eqnarray}
where ${1\over {\hat \epsilon}}\equiv {1\over \epsilon} (4 \pi)^\epsilon
\Gamma(1+\epsilon)$. 
We also define the following integrals,
\begin{eqnarray}
I_{1}(a) & = & \int_{0}^{1} dx \; \ln [ 1-ax(1-x)]
\\
I_{3}(a) & = & \int_{0}^{1} dx \; x(1-x) \ln [1-ax(1-x)]
\\
I_{4} (a,b) & = & \int_{0}^{1} dx \; \ln [1-x+ax-bx(1-x)]   
\end{eqnarray}

\section{One-Loop Contributions to Gauge Boson Self-Energies}\label{sec:one-loop}

The self-energies of the gauge bosons have the following structure
\begin{equation}
\Pi_{ij} (p^{2}) = g_{\mu\nu} \; \Pi_{ij}^{T} (p^{2}) 
+ p_{\mu} p_{\nu} \; \Pi_{ij}^{L} (p^{2}) \quad .
\end{equation}
Only the coefficient of the $g_{\mu\nu}$ term, the  
transverse part of the self-energy, contributes to the mass renormalization 
of the gauge boson. 
We calculate the one-loop contributions to the gauge boson self-energies in 
unitary gauge, in which the contributions from the non-physical particles 
vanish. The fermion contribution to 
the gauge boson self-energy is gauge invariant and finite. 
Our calculation manifests these properties; this serves 
as a cross-check of our result. In the bosonic sector, the total contribution 
is gauge dependent and is Ultra-Violet-divergent~\cite{Degrassi:1992ff}.  
Nevertheless, one can show that the contribution which is logarithmically 
enhanced by $\ln (M_{\Phi}/m_{Z})$ is gauge independent, using Eq.(7)-(9) 
of \cite{Degrassi:1992ff}.

\subsection{Contributions of a fermion loop}

The contribution due to the fermion loops to $\Pi^{XY}$, where 
$(XY)= \; (WW), \; (ZZ), \; (\gamma \gamma), \; (\gamma Z)$, is given by
\begin{eqnarray}
\Pi^{XY}(p^{2}) & = & -\frac{1}{16\pi^{2}} \biggl[
(c_{L}^{2}+c_{R}^{2}) \biggl( 2A_{0}(m_{2}^{2}) +
2m_{1}^{2}B_{0}(p^{2},m_{1}^{2},m_{2}^{2}) 
\\
& & \quad
-2p^{2}B_{1}(p^{2},m_{1}^{2},m_{2}^{2})
-4B_{22}(p^{2},m_{1}^{2},m_{2}^{2}) \biggr)
\nonumber\\
& & \quad -4
c_{L} c_{R} m_{1} m_{2} B_{0}(p^{2}, m_{1}^{2},m_{2}^{2}) 
\biggr] \quad .
\nonumber
\end{eqnarray}
where $m_{1}$ and $m_{2}$ are the masses of the loop fermion doublets.
At zero momentum transfer, this becomes,
\begin{eqnarray}
\Pi^{XY}(0) & = & 
-\frac{1}{16\pi^{2}} \biggl[ 
2(g_{V}^{2}+g_{A}^{2}) \biggl( \frac{1}{2} 
(m_{1}^{2} + m_{2}^{2}) + \frac{m_{1}^{4}}{m_{1}^{2}-m_{2}^{2}} 
\ln({Q^{2} \over m_{1}^{2}})
- \frac{m_{2}^{4}}{m_{1}^{2}-m_{2}^{2}} \ln({Q^{2} \over m_{2}^{2}})
\biggr)
\\
& & \quad 
-4 ( g_{V}^{2}-g_{A}^{2}) m_{1} m_{2} \biggl( 1 + \ln( {Q^{2} \over m_{2}^{2}}) 
+ \frac{m_{1}^{2}}{m_{1}^{2}-m_{2}^{2}} \ln({m_{2}^{2} \over m_{1}^{2}})
\biggr) \biggr] \quad .
\nonumber
\end{eqnarray}
Note that in the above expression the  contribution proportional to 
$1/\hat{\epsilon}$ has been subtracted. 
We define the following shorthand notations
\begin{eqnarray}
f_{1}(m_{1}^{2},m_{2}^{2}) & = & \frac{1}{2} 
(m_{1}^{2} + m_{2}^{2}) + \frac{m_{1}^{4}}{m_{1}^{2}-m_{2}^{2}} \ln( {Q^{2} \over m_{1}^{2}})
- \frac{m_{2}^{4}}{m_{1}^{2}-m_{2}^{2}} \ln( {Q^{2} \over m_{2}^{2}})
\\
f_{2}(m_{1}^{2},m_{2}^{2}) & = &
1 + \ln( {Q^{2} \over m_{2}^{2}} ) 
+ \frac{m_{1}^{2}}{m_{1}^{2}-m_{2}^{2}} \ln( {m_{2}^{2}\over m_{1}^{2}})\quad .
\end{eqnarray}


\subsection{Contributions of a pure scalar loop} 

The contribution of  scalar loops with 
two vectors and two scalars ($VVSS$) 
has no momentum dependence, and is given by
\begin{eqnarray}
\Pi^{XY}(p^{2}) = \Pi^{XY}(0) & = & \frac{1}{16\pi^{2}} 
C_{\scriptstyle XYS_{1}S_{2}} A_{0}(m^{2})
\\
& = & \frac{1}{16\pi^{2}} C_{\scriptstyle XYS_{1}S_{2}} m^{2} 
\biggl[ 1 + \ln ({Q^{2} \over m^{2}} ) + 
\frac{1}{\hat{\epsilon}} \biggr] \quad .
\nonumber
\end{eqnarray}
where $(XY) = (WW), \; (ZZ), \; (\gamma \gamma), \; (\gamma Z)$, and 
$m$ is the mass of the loop scalar fields.
Note there is an extra symmetry factor $1/2$ 
if the particle in the loop is neutral. 
The contribution of the scalar loops with one vector
and two scalars ($VSS$) 
is given by
\begin{equation}
\Pi^{XY}(p^{2}) = -\frac{4}{16\pi^{2}} C_{\scriptstyle XS_{1}S_{2}} 
C_{\scriptstyle YS_{1}S_{2}} \; 
B_{22}(p^{2},m_{1}^{2},m_{2}^{2})
\end{equation}
where $(XY) = (WW), \; (ZZ), \; (\gamma \gamma), \; (\gamma Z)$, 
and $m_{1}$ and $m_{2}$ are the masses of the loop scalar fields.
For $p^{2}=0$,
\begin{eqnarray}
\Pi^{XY}(0)
& = & -\frac{4}{16\pi^{2}} C_{\scriptstyle XS_{1}S_{2}} 
C_{\scriptstyle YS_{1}S_{2}} 
\; \biggl[ \, 
\frac{3}{8}(m_{1}^{2}+m_{2}^{2}) + \frac{1}{4(m_{1}^{2}-m_{2}^{2})}
\biggl( m_{1}^{4} \ln({Q^{2} \over m_{1}^{2}}) 
- m_{2}^{4} \ln({Q^{2}\over m_{2}^{2}}) \biggr) 
\\
& & \quad 
+ \, 
\biggl(\frac{m_{1}^{2}+m_{2}^{2}}{4}\biggr)
 \frac{1}{\hat{\epsilon}} \, \biggr]  \; .
 \nonumber
\end{eqnarray}
We define the shorthand notation
\begin{equation}
g_{1}(m_{1}^{2},m_{2}^{2}) = 
\frac{3}{8}(m_{1}^{2}+m_{2}^{2}) + \frac{1}{4(m_{1}^{2}-m_{2}^{2})}
\biggl[
m_{1}^{4} \ln({Q^{2}\over m_{1}^{2}}) 
- m_{2}^{4} \ln({Q^{2}\over m_{2}^{2}}) \biggr] \quad .
\end{equation}
In the limit $m_{1}^{2} = m_{2}^{2} = m^{2}$, this becomes
\begin{equation}
\Pi^{XY}(0)= -\frac{2}{16\pi^{2}} C_{\scriptstyle XS_{1}S_{2}} 
C_{\scriptstyle YS_{1}S_{2}}
m^{2} \biggl[ 1+\ln({Q^{2}\over m^{2}}) 
+ \frac{1}{\hat{\epsilon}} \biggr]\quad .
\end{equation}

\subsection{Contributions of a gauge boson-scalar loop}

The contribution of the gauge boson-scalar loops is given by
\begin{equation}
\Pi^{XX}(p^{2}) = \frac{1}{16\pi^{2}} C_{\scriptstyle XX^{\prime}S}^{2} 
\biggl[ B_{0}(p^{2},M_{X^{\prime}}^{2},m_{s}^{2}) 
- \frac{1}{M_{X^{\prime}}^{2}} 
B_{22}(p^{2},M_{X^{\prime}}^{2},m_{s}^{2}) \biggr]
\end{equation}
where $(XY) = (WW), \; (ZZ)$,  
$M_{X^{\prime}}$ is the mass of the loop gauge boson $X^{\prime}$, and 
$m_{s}$ is the mass of the loop scalar field.
For $p^{2}=0$, 
\begin{eqnarray}
\Pi^{XY}(0) & = & \frac{1}{16\pi^{2}} C_{\scriptstyle XX^{\prime}S}^{2} 
\biggl[
\frac{5}{8} - \frac{3}{8} \frac{m_{s}^{2}}{M_{X^{\prime}}^{2}} 
+ \frac{3}{4} (\frac{M_{X^{\prime}}^{2}}{M_{X^{\prime}}^{2}-m_{s}^{2}}) 
\ln({Q^{2}\over M_{X^{\prime}}^{2}}) 
\\
& & \quad
+ (\frac{m_{s}^{2}}{M_{X^{\prime}}^{2}-m_{s}^{2}})
(-1+\frac{m_{s}^{2}}{4M_{X^{\prime}}^{2}}) 
\ln({Q^{2}\over m_{s}^{2}}) 
+ (1-\frac{M_{X^{\prime}}^{2} + m_{s}^{2}}{4 M_{X^{\prime}}^{2}} )
\frac{1}{\hat{\epsilon}} \, \biggr]
\nonumber
\end{eqnarray}
The  contribution proportional to ${1\over 16\pi^2}
\ln(m_{s}^{2})$ is gauge invariant,
\begin{equation}\label{eq:g2}
g_{2}(m_{s}^{2},M_{X^{\prime}}^{2}) \equiv 
\biggl(\frac{m_{s}^{2}}{M_{X^{\prime}}^{2}-m_{s}^{2}}\biggr)
\biggl[(-1+\frac{m_{s}^{2}}{4M_{X^{\prime}}^{2}}) 
\ln({Q^{2}\over m_{s}^{2}})\biggr].
\end{equation}

\section{Gauge boson self-energies in the LLH model}

In our renormalization procedure, we need to calculate the following gauge 
boson self-energies,
$\Pi^{\gamma\gamma\prime}(0)$, $\Pi^{\gamma Z}(0)$, $\Pi^{\gamma Z}(M_{Z}^{2})$, 
$\Pi^{WW}(0)$ and $\Pi^{ZZ}(M_{Z}^{2})$. 
Below we summarize the full results for diagrams 
due to fermion and scalar loops. In our numerical results, we keep only the 
contributions which are enhanced by large logarithms, $\ln(M^2/Q^2)$,
where $M$ is a heavy mass scale and $Q$ is typically the weak
scale. The gauge independence  
in the bosonic sector can be retained by using the pinch technique or by using 
the background field formalism. This will be discussed in \cite{chen:2003lh2}.

\subsection{Contributions to $\Pi^{\gamma \gamma \prime}(0)$}

There are five diagrams that contribute to $\Pi^{\gamma\gamma}(0)$ 
in the LLH model.
These are loops having $(\overline{t}t)$, $(\overline{b}b)$, 
$(\overline{T}T)$, $(\Phi^{+} \Phi^{-})$, and $(\Phi^{++} \Phi^{--})$.
The total contribution to $\Pi^{\gamma\gamma\prime}(0)$ in the 
LLH model is
\begin{equation}
\Pi^{\gamma\gamma\prime}(0) = 
\frac{\alpha}{4\pi} \biggl[ \frac{5}{3} \ln \frac{Q^{2}}{M_{\Phi}^{2}}
+ \frac{16}{9} \ln\frac{Q^{2}}{m_{t}^{2}}
+ \frac{4}{9} \ln\frac{Q^{2}}{m_{b}^{2}}
+ \frac{16}{9} \ln\frac{Q^{2}}{m_{T}^{2}}
+ \frac{17}{3\hat{\epsilon}} \biggr] 
\end{equation}

\subsection{Contributions to $\Pi^{\gamma Z}(p^{2})$}

In the LLH model, there are six diagrams that contribute to 
$\Pi^{\gamma Z}(M_{Z}^{2})$. These are fermionic loops having 
$(\overline{t}t)$, $(\overline{T}T)$, 
the scalar loops due to $SSV$ couplings, $(\Phi^{+},\Phi^{-})$, 
$(\Phi^{++},\Phi^{--})$, and the $\Phi^{+}$ and $\Phi^{++}$ 
scalar loops due to $SSVV$ quartic couplings. The contributions to 
$\Pi^{\gamma Z}(M_{Z}^{2})$ due to the fermions are 
\begin{eqnarray}
\Pi^{\gamma Z}_{\scriptscriptstyle (\overline{t}t)}
(M_{Z}^{2}) & = & \frac{2\alpha}{\pi s_{W}c_{W}}
(\frac{1}{2}-\frac{4}{3}s_{W}^{2}) M_{Z}^{2}
\biggl[ \frac{1}{3}\biggl( 
\ln \frac{Q^{2}}{m_{t}^{2}}+\frac{1}{\hat{\epsilon}}) 
-2I_{3}(\frac{M_{Z}^{2}}{m_{t}^{2}} \biggr) \biggr]
\\
\Pi^{\gamma Z}_{\scriptscriptstyle (\overline{T}T)}
(M_{Z}^{2}) & = & -\frac{2\alpha}{\pi s_{W}c_{W}}
(\frac{4s_{W}^{2}}{3})(\frac{1}{2}-\frac{4}{3}s_{W}^{2}) M_{Z}^{2}
\\
&& \qquad \cdot
\biggl[ \frac{1}{3} \biggl(
\ln \frac{Q^{2}}{m_{T}^{2}}+\frac{1}{\hat{\epsilon}} \biggr) 
-2I_{3}(\frac{M_{Z}^{2}}{m_{T}^{2}}) \biggr] \nonumber
\end{eqnarray}
The sum of the contributions due to $SSV$ couplings is
\begin{eqnarray}\label{eq:pigzssv}
\Pi^{\gamma Z}(M_{Z}^{2}) & = & 
\frac{\alpha}{2\pi}
(\frac{s_{W}}{c_{W}})(5-\frac{2}{s_{W}^{2}})
\biggl[ \biggl( M_{\Phi}^{2}-\frac{1}{6}M_{Z}^{2} \biggr) 
\biggl( \ln \frac{Q^{2}}{M_{\Phi}^{2}} + \frac{1}{\hat{\epsilon}} \biggr)
\\
& & \quad
+ \biggl( \frac{1}{6}M_{Z}^{2}-\frac{2}{3}M_{\Phi}^{2} \biggr)
I_{1} \biggl( \frac{M_{Z}^{2}}{M_{\Phi}^{2}} \biggr)
+ M_{\Phi}^{2}-\frac{1}{9}M_{Z}^{2} \biggr]
\; . \nonumber
\end{eqnarray}
The sum of the contributions due to $SSVV$ couplings is
\begin{equation}\label{eq:pigzssvv}
\Pi^{\gamma Z}(M_{Z}^{2}) =  
-\frac{\alpha}{2\pi}
\biggl( \frac{s_{W}}{c_{W}} \biggr) \biggl( 5-\frac{2}{s_{W}^{2}} \biggr)
\biggl[ 
1+\ln \frac{Q^{2}}{M_{\Phi}^{2}} 
+ \frac{1}{\hat{\epsilon}} \biggr] M_{\Phi}^{2}
\end{equation}
The terms proportional to $M_{\Phi}^{2}$ and $M_{\Phi}^{2}\ln (Q^{2}/M_{\Phi}^{2})$ 
in Eq.(\ref{eq:pigzssv}) and (\ref{eq:pigzssvv}) cancel among them-selves. 
The total contribution to $\Pi^{\gamma Z}(M_{Z}^{2})$ is thus given by, 
to order $\mathcal{O}(1/16\pi^{2})$,
\begin{eqnarray}
\Pi^{\gamma Z}(M_{Z}^{2}) & = &
\frac{2\alpha}{\pi s_{\theta}c_{\theta}}
\biggl( \frac{1}{2}-\frac{4}{3}s_{\theta}^{2} \biggr) M_{Z}^{2}
\\
& & \quad \cdot \biggl[
\frac{1}{3} \ln (\frac{Q^{2}}{m_{t}^{2}}) - 2 I_{3}(\frac{M_{Z}^{2}}{m_{t}^{2}})
-\frac{4s_{\theta}^{2}}{3} \biggl(
\frac{1}{3}\ln (\frac{Q^{2}}{M_{T}^{2}}) - 2 I_{3}(\frac{M_{Z}^{2}}{M_{T}^{2}} \biggr)
\biggl]
\nonumber\\
&& + \frac{\alpha s_{\theta}}{2 \pi c_{\theta}} (5-\frac{2}{s_{\theta}^{2}})
\biggl[
-\frac{2}{3}M_{\Phi}^{2}
I_{1}(\frac{M_{Z}^{2}}{M_{\Phi}^{2}})
-\frac{1}{6}M_{Z}^{2} \biggl( \ln(\frac{Q^{2}}{M_{\Phi}^{2}}) 
-I_{1}(\frac{M_{Z}^{2}}{M_{\Phi}^{2}}) +\frac{2}{3} \biggr) \biggr] \; .
\nonumber
\end{eqnarray}

For $p^{2}=0$, it can be easily checked that the total fermionic 
contribution and the total scalar contribution to $\Pi^{\gamma Z}(0)$ 
vanish individually. Thus 
\begin{equation}
\Pi^{\gamma Z}(0)=0,
\end{equation}
as expected in the unitary gauge.

\subsection{Contributions to $\Pi^{WW}(0)$}

The full list of contributions of fermion loops to $\Pi^{WW}(0)$ 
is given as follows,
\begin{eqnarray}
\Pi^{WW}_{(\overline{t}b)}(0) & = &
-\frac{1}{16\pi^{2}} \frac{g^{2}}{2} f_{1}(m_{t}^{2},m_{b}^{2})
\nonumber
\\
\Pi^{WW}_{(\overline{T}b)}(0) & = &
-\frac{1}{16\pi^{2}} \frac{g^{2}}{2} (\frac{v}{f})^{2} x_{L}^{2} 
\, f_{1}(m_{T}^{2},m_{b}^{2}) \quad .
\end{eqnarray}
The sum of the fermionic contributions to $\Pi^{WW}(0)$ is thus given by, 
to order $\mathcal{O}(1/16\pi^{2})$,
\begin{eqnarray}
\Pi^{WW}_{f}(0) & = & 
-\frac{\alpha}{8\pi s_{\theta}^{2}} \biggl[
f_{1}(m_{t}^{2},m_{b}^{2}) + x_{L}^{2} (\frac{1}{\sqrt{2}G_{\mu}f^{2}}) M_{T}^{2} 
\biggl( \frac{1}{2} + \frac{M_{T}^{2}}{M_{T}^{2}-m_{b}^{2}}
\ln(\frac{Q^{2}}{M_{T}^{2}}) \biggr) \; \biggr]. 
\end{eqnarray}
where $M_{T}$ in the above equation is replaced by its leading order term, $M_{T}^{2} 
\rightarrow \frac{\sqrt{2}G_{\mu}m_{t}^{2}}{x_{L}(1-x_{L})} f^{2}$.
The full list
 of contributions of scalar loops to $\Pi^{WW}(0)$ is given as follows,
\begin{eqnarray}
\Pi^{WW}_{(s)}(0) & = & \frac{1}{16\pi^{2}} g^{2} \biggl[
\frac{1}{4} m_{H}^{2} \biggl( 1 + ln (\frac{Q^{2}}{m_{H}^{2}} ) \biggr) 
+ 4 M_{\Phi}^{2} \biggl( 1 + ln (\frac{Q^{2}}{M_{\Phi}^{2}} ) \biggr) \biggr]
\\ 
&& 
\biggl[ 
\mbox{\small s = sum of} \; {\scriptstyle h, \, \Phi^{0}, \, \Phi^{P}, \, 
\Phi^{+}, \, \Phi^{++}} \biggr]
\nonumber\\
\Pi^{WW}_{(s_{1}s_{2})}(0) & = & -\frac{4}{16\pi^{2}} g^{2} \biggl[
\frac{(\sqrt{2}s_{0}-s_{+})^{2}}{4} g_{1}(m_{H}^{2},M_{\Phi}^{2})
+ 2 g_{1}(M_{\Phi}^{2},M_{\Phi}^{2}) \biggr]
\\
& = & \frac{4}{16\pi^{2}} g^{2} \biggl[
\frac{(\sqrt{2}s_{0}-s_{+})^{2}}{4} g_{1}(m_{H}^{2},M_{\Phi}^{2})
+ M_{\Phi}^{2} \biggl( 1 + ln (\frac{Q^{2}}{M_{\Phi}^{2}} ) \biggr) \biggr]
\nonumber\\
& & \qquad \biggl[ 
{\scriptstyle (s_{1}, s_{2}) = \; \mbox{\small sum of} 
(H, \Phi^{-}), \, (\Phi^{0},\Phi^{-}), \, 
(\Phi^{P},\Phi^{-}), \, (\Phi^{+},\Phi^{--})
\biggr] }
\nonumber \quad .
\end{eqnarray}
Note that the contribution of the triplet components 
$(\Phi^{0},\Phi^{P},\Phi^{+},\Phi^{++})$ to $\Pi^{WW}_{(s)}(0)$ 
cancels {\it exactly} the contribution of 
$(\Phi^{0},\Phi^{-}), \, (\Phi^{P},\Phi^{-}), \, (\Phi^{+},\Phi^{--})$ to 
$\Pi^{WW}_{(s_{1}s_{2})}(0)$. This prevents the appearance 
of contributions proportional 
to $M_{\Phi}^{2}$ and $M_{\Phi}^{2} ln(\frac{Q^{2}}{M_{\Phi}^{2}}).$
To order $\mathcal{O}(1/16\pi^{2})$, the sum of the contributions due to pure 
scalar loops is
\begin{eqnarray}
\Pi^{WW}_{s}(0) & = &
\frac{\alpha}{4\pi s_{\theta}^{2}} \biggl[
\frac{1}{4}m_{H}^{2} \biggl( 1 + \ln(\frac{Q^{2}}{m_{H}^{2}}) \biggr)
\\
& & \quad 
- 4 \sqrt{2} G_{\mu} v^{\prime 2}  M_{\Phi}^{2}
\biggl( \frac{3}{8} + \frac{M_{\Phi}^{2}}{4(M_{\Phi}^{2}-m_{H}^{2})}
\ln (\frac{Q^{2}}{M_{\Phi}^{2}}) \biggr)
\; \biggr].
\end{eqnarray}

The complete list of contributions 
proportional to $\ln(m_{s}^{2})$ to $\Pi^{WW}(0)$ from scalar-gauge boson loops 
is,
\begin{eqnarray}
\Pi^{WW}(0) & = & \frac{1}{16\pi^{2}} \; \{ 
C_{{\scriptscriptstyle W_{L}W_{L}h}}^{2} g_{2}(m_{H}^{2},M_{W_{L}}^{2})
+ C_{\scriptscriptstyle W_{L}W_{H}h}^{2} g_{2}(m_{H}^{2},M_{W_{H}}^{2})
\nonumber\\
&& + C_{\scriptscriptstyle W_{L}W_{L}\Phi^{0}}^{2} g_{2}(M_{\Phi}^{2},M_{W_{L}}^{2})
+ C_{\scriptscriptstyle W_{L}W_{H}\Phi^{0}}^{2} g_{2}(M_{\Phi}^{2},M_{W_{H}}^{2})
\nonumber\\
&& + C_{\scriptscriptstyle W_{L}Z_{L}\Phi^{-}}^{2} g_{2}(M_{\Phi}^{2},M_{Z}^{2})
+ C_{\scriptscriptstyle W_{L}Z_{H}\Phi^{-}}^{2} g_{2}(M_{\Phi}^{2},M_{Z_{H}}^{2})
\nonumber\\
& & + C_{\scriptscriptstyle W_{L}A_{H}\Phi^{-}}^{2} g_{2}(M_{\Phi}^{2},M_{A_{H}}^{2})
+ C_{\scriptscriptstyle W_{L}W_{L}\Phi^{--}}^{2} g_{2}(M_{\Phi}^{2},M_{W_{L}}^{2})
\nonumber\\
&& + C_{\scriptscriptstyle W_{L}W_{H}\Phi^{--}}^{2} g_{2}(M_{\Phi}^{2},M_{W_{H}}^{2}) \}
\end{eqnarray}
where the gauge coupling constants of the scalar fields are summarized in Table \ref{XXS}.
To order $\mathcal{O}(1/16\pi^{2})$, the sum of the contributions due to scalar-gauge-boson 
loops is,
\begin{eqnarray}
\Pi^{WW}_{sv}(0) & = & 
\frac{\alpha^{2}}{\sqrt{2}G_{\mu}s_{\theta}^{4}}
\biggl[
\frac{1}{4} \biggl( \frac{m_{H}^{2}}{M_{W_{L}}^{2}-m_{H}^{2}} \biggr)
\biggl(-1+\frac{m_{H}^{2}}{M_{W_{L}}^{2}} \biggr) \ln(\frac{Q^{2}}{m_{H}^{2}})
\nonumber\\
& & \quad + \frac{\sqrt{2}G_{\mu}}{c_{\theta}^{2}}v^{\prime 2}
\biggl( \frac{M_{\Phi}^{2}}{M_{Z}^{2}-M_{\Phi}^{2}} \biggr) 
\biggl(\frac{M_{\Phi}^{2}}{M_{Z}^{2}} \biggr)
\ln (\frac{Q^{2}}{M_{\Phi}^{2}})
\nonumber\\
& & \quad 
+ 4\sqrt{2}G_{\mu} v^{\prime 2} 
\biggl( \frac{M_{\Phi}^{2}}{M_{W_{L}}^{2}-M_{\Phi}^{2}} \biggr)
\biggl(\frac{M_{\Phi}^{2}}{M_{W_{L}}^{2}} \biggr)
\ln (\frac{Q^{2}}{M_{\Phi}^{2}})
\biggr].
\end{eqnarray}

\subsection{Contributions to $\Pi^{ZZ}(M_{Z}^{2})$}

The complete list of fermionic contributions to the self-energy function 
$\Pi^{ZZ}(p^{2})$ are summarized below.
\begin{eqnarray}
\Pi^{ZZ}_{\scriptscriptstyle (\overline{T}t)} (M_{Z}^{2}) & = &
-\frac{1}{16\pi^{2}}(\frac{g x_{L}}{2c_{W}})^{2} \frac{v^{2}}{f^{2}} 
\biggl[
-\frac{1}{3M_{Z}^{2}}(M_{T}^{2}-m_{t}^{2})^{2} 
+ \frac{2}{9}M_{Z}^{2} 
\\
& & \quad + \frac{1}{6M_{Z}^{2}}
\biggl[-M_{T}^{4} + m_{t}^{4} - 2 M_{Z}^{4} + M_{Z}^{2}\biggl(5M_{T}^{2}+m_{t}^{2}\biggr)\biggr]
\ln \frac{Q^{2}}{M_{T}^{2}}
\nonumber\\
& & \quad + \frac{1}{6M_{Z}^{2}}
\biggl[-m_{t}^{4} + M_{T}^{4} - 2 M_{Z}^{4} + M_{Z}^{2}\biggl(5m_{t}^{2}+M_{T}^{2}\biggr)\biggr]
\ln \frac{Q^{2}}{m_{t}^{2}}
\nonumber\\
&&\quad
-\frac{1}{6}\biggl[m_{t}^{2} + M_{T}^{2}-2M_{Z}^{2}
+\frac{(m_{t}^{2}-M_{T}^{2})^{2}}{M_{Z}^{2}}\biggr]
\biggl(I_{4}(\frac{M_{T}^{2}}{m_{t}^{2}},\frac{M_{Z}^{2}}{m_{t}^{2}})
+I_{4}(\frac{M_{t}^{2}}{M_{T}^{2}},\frac{M_{Z}^{2}}{M_{T}^{2}})\biggr)
\nonumber\\
& & \quad
+\frac{1}{\hat{\epsilon}}\biggl[M_{T}^{2}+m_{t}^{2}-\frac{2}{3}M_{Z}^{2}\biggr] \; \biggr]
\nonumber \\
\Pi^{ZZ}_{\scriptscriptstyle (\overline{t}T)} (M_{Z}^{2}) & = &
\Pi^{ZZ}_{\scriptscriptstyle (\overline{T}t)} (M_{Z}^{2})
\\
\Pi^{ZZ}_{\scriptscriptstyle (\overline{t}t)} (M_{Z}^{2}) & = &
-\frac{1}{16\pi^{2}} (\frac{g}{2c_{W}})^{2} 
\biggl[ 2 
\biggl((\frac{1}{2}-\frac{4}{3}s_{W}^{2})^{2} + \frac{1}{4}\biggr) 
h_{1}(m_{t}^{2})
\\
& & \quad 
-\frac{16}{3}s_{W}^{2}(1-\frac{4}{3}s_{W}^{2}) h_{2}(m_{t}^{2})
\nonumber\\
& & \quad + \frac{1}{\hat{\epsilon}} 
\biggl[ 2 \biggl(\frac{1}{2}-\frac{4}{3}s_{W}^{2})^{2} + \frac{1}{4} \biggr)
(2m_{t}^{2}-\frac{2}{3}M_{Z}^{2}) 
+ \frac{16}{3}s_{W}^{2}(1-\frac{4}{3}s_{W}^{2}) m_{t}^{2} \biggr] \; \biggr]
\nonumber\\
\Pi^{ZZ}_{\scriptscriptstyle (\overline{b}b)} (M_{Z}^{2}) & = &
-\frac{1}{16\pi^{2}} (\frac{g}{2c_{W}})^{2} 
\biggl[ 2 \biggl(
(\frac{1}{2}-\frac{2}{3}s_{W}^{2})^{2} + \frac{1}{4} \biggr) h_{1}(m_{b}^{2})
\\
& & \quad 
-\frac{8}{3}s_{W}^{2}(1-\frac{2}{3}s_{W}^{2}) h_{2}(m_{b}^{2})
\nonumber\\
& & \quad + \frac{1}{\hat{\epsilon}} 
\biggl[ 2 \biggl( 
(\frac{1}{2}-\frac{2}{3}s_{W}^{2})^{2} + \frac{1}{4} \biggr)
(2m_{b}^{2}-\frac{2}{3}M_{Z}^{2}) 
+ \frac{8}{3}s_{W}^{2}(1-\frac{2}{3}s_{W}^{2}) m_{b}^{2} \biggr] \; \biggr]
\nonumber\\
\Pi^{ZZ}_{\scriptscriptstyle (\overline{T}T)} (M_{Z}^{2}) & = &
-\frac{1}{16\pi^{2}} (\frac{2s_{W}^{2}g}{3c_{W}})^{2}
\biggl[ -\frac{4}{3}M_{Z}^{2}
\biggl( \ln \frac{Q^{2}}{M_{T}^{2}}+\frac{1}{\hat{\epsilon}} \biggr)
+ \frac{4}{9}M_{Z}^{2} 
\\
& & \quad + \biggl( \frac{4}{3}M_{Z}^{2}+\frac{8}{3}M_{T}^{2} \biggr)
I_{1}(\frac{M_{Z}^{2}}{M_{T}^{2}}) \biggr] \; ,\nonumber
\end{eqnarray}
where $h_{1}(m^{2})$ and $h_{2}(m^{2})$ are defined as
\begin{eqnarray}
h_{1}(m^{2}) & = & 
(2m^{2}-\frac{2}{3}M_{Z}^{2})
\ln \frac{Q^{2}}{m^{2}} + \frac{2}{9} M_{Z}^{2} 
+\frac{2}{3}(M_{Z}^{2}-m^{2})I_{1}(\frac{M_{Z}^{2}}{m^{2}})
\\
h_{2}(m^{2}) & = &
m^{2}I_{1}(\frac{M_{Z}^{2}}{m^{2}})
-m^{2} \ln \frac{Q^{2}}{m^{2}} \; .
\end{eqnarray}
To order $\mathcal{O}(1/16 \pi^{2})$, the sum of the fermionic contributions to 
$\Pi^{ZZ}(M_{Z}^{2})$ is,
\begin{eqnarray}
\Pi^{ZZ}_{f}(M_{Z}^{2}) & = &
-\frac{\alpha}{8\pi s_{\theta}^{2}c_{\theta}^{2}}
\nonumber\\
& & \cdot
\biggl[ \quad
(\frac{x_{L}^{2}}{\sqrt{2}G_{\mu}f^{2}}) M_{T}^{2} \biggl[
\frac{2m_{t}^{2}}{3M_{Z}^{2}} + \frac{5}{6}\ln(\frac{Q^{2}}{M_{T}^{2}})
+ \frac{1}{6} \ln (\frac{Q^{2}}{m_{t}^{2}})
\nonumber\\
& & \qquad \qquad 
- \frac{1}{6} \biggl( 1-\frac{2m_{t}^{2}}{M_{Z}^{2}} \biggr)
\biggl( I_{4} (\frac{M_{T}^{2}}{m_{t}^{2}},\frac{M_{Z}^{2}}{m_{t}^{2}}) 
+I_{4} (\frac{m_{t}^{2}}{M_{T}^{2}},\frac{M_{Z}^{2}}{M_{T}^{2}}) \biggr) \; \biggr]
\nonumber\\
&& \qquad - (\frac{x_{L}^{2}}{3\sqrt{2}G_{\mu}f^{2}}) \biggl( 
1 + \frac{\Delta s_{\theta}^{2}}{c_{\theta}^{2}}
- \frac{\Delta s_{\theta}^{2}}{s_{\theta}^{2}} -\frac{1}{4\sqrt{2}G_{\mu}f^{2}}
-4\frac{v^{\prime 2}}{f^{2}} \biggr)
\frac{M_{T}^{4}}{M_{Z}^{2}}
\biggl[ 1 
\nonumber\\
& & \qquad \qquad + \frac{1}{2} 
\biggl( \ln (\frac{m_{t}^{2}}{M_{T}^{2}})
+ I_{4}(\frac{M_{T}^{2}}{m_{t}^{2}},\frac{M_{Z}^{2}}{m_{t}^{2}}) 
+ I_{4}(\frac{m_{t}^{2}}{M_{T}^{2}},\frac{M_{Z}^{2}}{M_{T}^{2}}) \biggr)
\biggr]
\nonumber\\
& & \qquad
+ \biggl( (\frac{1}{2}-\frac{4}{3}s_{\theta}^{2})^{2} + \frac{1}{4} \biggr) 
h_{1}(m_{t}^{2}) - \frac{8}{3} s_{\theta}^{2} (1-\frac{4}{3}s_{\theta}^{2}) h_{2}(m_{t}^{2})
\nonumber\\
& & \qquad
+  \biggl( (\frac{1}{2}-\frac{2}{3}s_{\theta}^{2})^{2} + \frac{1}{4} \biggr) 
h_{1}(m_{b}^{2}) - \frac{4}{3} s_{\theta}^{2} (1-\frac{2}{3}s_{\theta}^{2}) h_{2}(m_{b}^{2})
\nonumber\\
& & \qquad
+ \frac{8 }{9} s_{\theta}^{4} M_{Z}^{2} \biggl[
-\frac{4}{3} \ln (\frac{Q^{2}}{M_{T}^{2}}) + \frac{4}{9} + \frac{4}{3}
I_{1}(\frac{M_{Z}^{2}}{M_{T}^{2}}) \biggr] \quad \biggr]
\end{eqnarray}

The sum of the scalar contributions due to VVSS quartic couplings is
\begin{eqnarray}
\Pi^{ZZ}_{(s)}(M_{Z}^{2}) & = & \frac{1}{16\pi^{2}} \frac{g^{2}}{c_{W}^{2}} 
\biggl[
\frac{1}{4} m_{H}^{2} \biggl( 1 + \ln (\frac{Q^{2}}{m_{H}^{2}} ) \biggr) 
\nonumber\\
& & + 2 \biggl(
1 + s_{W}^{4} + (1-2s_{W}^{2})^{2} \biggr) M_{\Phi}^{2} 
\biggl( 1 + \ln (\frac{Q^{2}}{M_{\Phi}^{2}}) \biggr) \; \biggr] 
\\ 
&& 
\biggl[
\mbox{\small s = sum of} \; {\scriptstyle h, \, \Phi^{0}, \, \Phi^{P}, \, 
\Phi^{+}, \, \Phi^{++}} \biggr]\;.
\nonumber
\end{eqnarray}

The sum of scalar contributions due to $(\Phi^{0}\Phi^{P})$, 
$(\Phi^{+}\Phi^{-})$ and $(\Phi^{++}\Phi^{--})$ loops, is given by, 
\begin{eqnarray}
\Pi^{ZZ}_{(s_{1}s_{2})}(M_{Z}^{2}) & = &
-\frac{1}{16\pi^{2}} (\frac{2g^{2}}{3c_{W}^{2}}) 
\biggl( 1+s_{W}^{4} +(1-2s_{W}^{2})^{2} \biggr)
\biggl[ 3M_{\Phi}^{2} - \frac{1}{3} M_{Z}^{2} 
\\
& & \qquad
+ \biggl( 3M_{\Phi}^{2} - \frac{1}{2} M_{Z}^{2} \biggr)
\biggl(
\ln \frac{Q^{2}}{M_{\Phi}^{2}} + \frac{1}{\hat{\epsilon}} \biggr) 
+\biggl(
\frac{1}{2}M_{Z}^{2}-2M_{\Phi}^{2} \biggr)
I_{1}(\frac{M_{Z}^{2}}{M_{\Phi}^{2}}) \; \biggr].\nonumber
\end{eqnarray}
The contributions proportional to $M_{\Phi}^{2}$ and 
$M_{\Phi}^{2}\ln \frac{Q^{2}}{M_{\Phi}^{2}}$ due to VVSS quartic couplings 
cancel exactly those due to VSS couplings. Thus there is no contribution 
proportional to $M_{\Phi}^{2}$ and $M_{\Phi}^{2}\ln (\frac{Q^{2}}{M_{\Phi}^{2}})$
due to pure scalar loops.
For the contribution due to $(H\Phi^{P})$ loop, we have
\begin{eqnarray}
\Pi^{ZZ}_{(H\Phi^{P})}(M_{Z}^{2}) & = &
-\frac{1}{16\pi^{2}} (\frac{g}{c_{W}})^{2} (s_{p}-2s_{0})^{2} \frac{1}{12}
\nonumber\\
& & \quad \cdot
\biggl[ \frac{1}{2} 
\biggl( 3 M_{\Phi}^{2} + 3 m_{H}^{2} - M_{Z}^{2} \biggr)
\biggl( \frac{1}{\hat{\epsilon}} + 
\ln (\frac{Q^{4}}{M_{\Phi}^{2}m_{H}^{2}} ) \biggr)
\nonumber\\
& & \qquad 
+ \frac{1}{2M_{Z}^{2}} \biggl(
M_{\Phi}^{4} - m_{H}^{4} + M_{Z}^{2}(M_{\Phi}^{2}-m_{H}^{2})
\biggr) \ln (\frac{M_{\Phi}^{2}}{m_{H}^{2}})
\nonumber\\
& & \qquad
+ \frac{1}{2M_{Z}^{2}} \biggl( 
M_{\Phi}^{4} + ( m_{H}^{2} - M_{Z}^{2} )^{2}
-2 M_{\Phi}^{2} ( m_{H}^{2} + M_{Z}^{2} ) \; \biggr) 
\nonumber\\
& & \qquad \qquad \qquad \cdot
\biggl( I_{4}(\frac{m_{H}^{2}}{M_{\Phi}^{2}}, \frac{M_{Z}^{2}}{M_{\Phi}^{2}})
+ I_{4}(\frac{M_{\Phi}^{2}}{m_{H}^{2}}, \frac{M_{\Phi}^{2}}{m_{H}^{2}})
\biggr)
\nonumber\\
& & \qquad
+ \frac{1}{3M_{Z}^{2}} \biggl(
3M_{\Phi}^{4} + M_{\Phi}^{2} (9M_{Z}^{2}-6m_{H}^{2}) + 3m_{H}^{4} 
\nonumber\\
& & \qquad \qquad \qquad \qquad 
-2M_{Z}^{4} + 9m_{H}^{2}M_{Z}^{2} \biggr) \; 
\biggr] \; . \quad
\end{eqnarray}
In terms of the input parameters, the sum of the contributions to $\Pi^{ZZ}(M_{Z}^{2})$ 
due to pure scalar loop to $\mathcal{O}(1/16\pi^{2})$ is,
\begin{eqnarray}
\Pi^{ZZ}_{s}(M_{Z}^{2}) & = &
\frac{\alpha}{4\pi s_{\theta}^{2}c_{\theta}^{2}}
\biggl[
\frac{1}{4}m_{H}^{2} \biggl( 1 + \ln (\frac{Q^{2}}{m_{H}^{2}})  \biggr)
\nonumber\\
& & \qquad
+ \biggl( 1 + s_{\theta}^{4} + (1-2s_{\theta}^{2})^{2} \biggr)
\biggl[ \frac{2}{3} M_{Z}^{2} \biggl(
\frac{1}{3} + \frac{1}{2} \ln(\frac{Q^{2}}{M_{\Phi}^{2}}) 
-\frac{1}{2} I_{1} \biggl( \frac{M_{Z}^{2}}{M_{\Phi}^{2}} \biggr) \, \biggl)
+ \frac{4}{3} M_{\Phi}^{2} I_{1}(\frac{M_{Z}^{2}}{M_{\Phi}^{2}}) \biggr] 
\nonumber\\
& & \qquad - \frac{\sqrt{2}G_{\mu}v^{\prime 2}}{12}  M_{\Phi}^{2} \biggl[
\frac{3}{2} \ln (\frac{Q^{4}}{m_{H}^{2}M_{\Phi}^{2}}) 
+ \frac{1}{2} \ln (\frac{M_{\Phi}^{2}}{m_{H}^{2}})
+ \biggl( 3 - \frac{2m_{H}^{2}}{M_{Z}^{2}} \biggr)
\nonumber\\
& & \qquad \qquad \qquad
-\frac{m_{H}^{2}}{M_{Z}^{2}} \biggl(
I_{4}(\frac{m_{H}^{2}}{M_{\Phi}^{2}},\frac{M_{Z}^{2}}{M_{\Phi}^{2}})
+ I_{4}(\frac{M_{\Phi}^{2}}{m_{H}^{2}},\frac{M_{Z}^{2}}{m_{H}^{2}})
\biggr) \biggr] 
\nonumber\\
 & & \qquad - \frac{\sqrt{2}G_{\mu}v^{\prime 2}}{24} 
\biggl( 1 - \frac{\Delta s_{\theta}^{2}}{s_{\theta}^{2}} + 
\frac{\Delta s_{\theta}^{2}}{c_{\theta}^{2}} -\frac{1}{4\sqrt{2}G_{\mu}f^{2}} 
-\frac{4v^{\prime 2}}{f^{2}} \biggr) 
\frac{M_{\Phi}^{4}}{M_{Z}^{2}}
\nonumber\\
&& \qquad \qquad \qquad \cdot
\biggl[ 2 + \ln (\frac{M_{\Phi}^{2}}{m_{H}^{2}}) 
+ I_{4}(\frac{m_{H}^{2}}{M_{\Phi}^{2}}, \frac{M_{Z}^{2}}{M_{\Phi}^{2}}) 
+ I_{4} (\frac{M_{\Phi}^{2}}{m_{H}^{2}},\frac{M_{Z}^{2}}{m_{H}^{2}})
\biggr] \; \biggr].
\end{eqnarray}
The contributions to $\Pi^{ZZ}(M_{Z}^{2})$ due to scalar-gauge-boson loops
have the following form
\begin{eqnarray}
\Pi^{ZZ}_{\scriptscriptstyle (VS)}(M_{Z}^{2}) & = & 
\frac{1}{16\pi^{2}} C_{XX^{\prime}S}^{2} 
\nonumber\\
& & \quad \cdot \biggl[ \;
\biggl( 1-\frac{1}{12M_{X^\prime}^{2}} \biggl( 3M_{X^\prime}^{2} 
+ 3 m_{S}^{2} - M_{Z}^{2} \biggr) \, \biggr) 
\biggl( \frac{1}{2} \ln \biggl( \frac{Q^{4}}{M_{X^\prime}^{2}m_{S}^{2}} \biggr) 
+ \frac{1}{\hat{\epsilon}} \biggr)
\nonumber\\
& & \qquad 
-\biggl( \frac{1}{2} + \frac{1}{24M_{Z}^{2}M_{X^\prime}^{2}} 
\biggl( M_{X^\prime}^{4} + (m_{S}^{2} - M_{Z}^{2})^{2}
- 2 M_{X^\prime}^{2} \biggl( m_{S}^{2} + M_{Z}^{2} \biggr) \; \biggr) \; \biggr)
\nonumber\\
& & \qquad \qquad 
\cdot \biggl( 
I_{4} \biggl( \frac{m_{S}^{2}}{M_{X^\prime}^{2}}, 
\frac{M_{Z}^{2}}{M_{X^\prime}^{2}} \biggr)  
+ I_{4} \biggl( \frac{M_{X^\prime}^{2}}{m_{S}^{2}}, 
\frac{M_{Z}^{2}}{m_{s}^{2}} \biggr) \,
\biggr) 
\nonumber\\
& & \qquad 
- \frac{1}{24M_{Z}^{2}M_{X^\prime}^{2}} 
\biggl( M_{X^\prime}^{4} - m_{S}^{4} 
+ M_{Z}^{2}(M_{X^{\prime}}^{2}-m_{S}^{2}) \biggr) 
\ln \biggl( \frac{M_{X^{\prime}}^{2}}{m_{S}^{2}} \biggr)
\nonumber\\
& & \qquad 
- \frac{1}{36M_{Z}^{2}M_{X^\prime}^{2}}
\biggl( 3M_{X^{\prime}}^{4} + M_{X^{\prime}}^{2} (9M_{Z}^{2}-6m_{S}^{2})
+ 3m_{S}^{4} - 2 M_{Z}^{4} + 9 m_{S}^{2} M_{Z}^{2} \biggr)
 \;\biggr] \; . 
\nonumber\\
\end{eqnarray}
where $M_{X^{\prime}}$ is the mass of the loop gauge boson and $m_{S}$ is the mass 
of the loop scalar field.
The contribution proportional to $\ln(m_{s}^{2})/16\pi^{2}$ is
\begin{eqnarray}
g_{3}(m_{s}^{2},M_{X^\prime}^{2}) & \equiv &
\frac{1}{2} \biggl[
1-\frac{1}{12 M_{X^\prime}^{2}}
\biggl( 3m_{s}^{4}+3M_{X^{\prime}}^{2}-M_{Z}^{2} \biggr) \; \biggr]
\ln (\frac{Q^{2}}{m_{s}^{2}}) \; .
\end{eqnarray}
Using this notation, the total contribution to $\Pi^{ZZ}(M_{Z}^{2})$ 
proportional to $\ln(m_{s}^{2})$ 
from scalar-gauge boson loops, is
\begin{eqnarray}
\Pi^{ZZ}_{\scriptscriptstyle (VS)} (M_{Z}^{2}) & = & 
\frac{1}{16\pi^{2}} 
\; \biggl[ 
C_{{\scriptscriptstyle Z_{L}Z_{L}h}}^{2} g_{3}(m_{H}^{2},M_{Z}^{2})
+ C_{\scriptscriptstyle Z_{L}A_{H}h}^{2} g_{3}(m_{H}^{2},M_{A_{H}}^{2})
\\
&& + C_{\scriptscriptstyle Z_{L}Z_{L}\Phi^{0}}^{2} g_{3}(M_{\Phi}^{2},M_{Z}^{2})
+ C_{\scriptscriptstyle Z_{L}Z_{H}\Phi^{0}}^{2} g_{3}(M_{\Phi}^{2},M_{Z_{H}}^{2})
\nonumber\\
&& + C_{\scriptscriptstyle Z_{L}Z_{H}H}^{2} g_{3}(m_{H}^{2},M_{Z_{H}}^{2})
+ C_{\scriptscriptstyle Z_{L}A_{H}\Phi^{0}}^{2} g_{3}(M_{\Phi}^{2},M_{A_{H}}^{2})
\nonumber\\
& & + C_{\scriptscriptstyle Z_{L}W_{L}\Phi^{-}}^{2} g_{3}(M_{\Phi}^{2},M_{W_{L}}^{2})
+ C_{\scriptscriptstyle Z_{L}W_{H}\Phi^{-}}^{2} g_{3}(M_{\Phi}^{2},M_{W_{H}}^{2})
\biggr]
\nonumber
\end{eqnarray}
where the gauge coupling constants of the scalar fields are summarized in 
Table \ref{XXS}. 
Expanding the coupling constants and masses in terms of the input parameters,  
to $\mathcal{O}(1/16\pi^{2})$, 
$\Pi^{ZZ}_{\scriptscriptstyle (VS)} (M_{Z}^{2})$ is given by
\begin{eqnarray}
\Pi^{ZZ}_{\scriptscriptstyle (VS)} (M_{Z}^{2}) & = & 
\frac{\alpha^{2}}{8\sqrt{2}G_{\mu}s_{\theta}^{4}c_{\theta}^{4}}
\nonumber\\
& & \cdot \biggl[ \quad
\biggl( \frac{5}{6} - \frac{m_{H}^{2}}{4M_{Z}^{2}} 
+ \frac{3(c^{\prime 2} - s^{\prime2})^{2}}{16s^{\prime2} c^{\prime2}}
+ \frac{3(c^{2} - s^{2})^{2}}{16 s^{2} c^{2}}
\; \biggr)
\ln \biggl( \frac{Q^{2}}{m_{H}^{2}} \biggr) 
\nonumber\\
& & \qquad 
+ \biggl[ \;  8 \sqrt{2} G_{\mu} v^{\prime 2} \biggl(
\frac{5}{6}-\frac{M_{\Phi}^{2}}{4M_{Z}^{2}} \biggr) 
\nonumber\\
& & \qquad \qquad 
+ \frac{(c^{2} - s^{2})^{2}}{4 s^{2} c^{2}}
\biggl( \frac{3}{4} - \frac{M_{\Phi}^{2}}{4M_{Z_{H}}^{2}} \biggr)
\nonumber\\
& & \qquad \qquad
+ \frac{\sqrt{2} G_{\mu} (c^{\prime 2} - s^{\prime 2})^{2}v^{\prime 2}}
{4 s^{\prime 2} c^{\prime 2}}
\biggl( \frac{3}{4} - \frac{M_{\Phi}^{2}}{4M_{A_{H}}^{2}} \biggr)
\; \biggr]
\ln \biggl( \frac{Q^{2}}{M_{\Phi}^{2}} \biggr) \; \biggr]
\nonumber\\
\end{eqnarray}

\pagebreak

\begin{figure}
{\center
\includegraphics[scale=0.75]{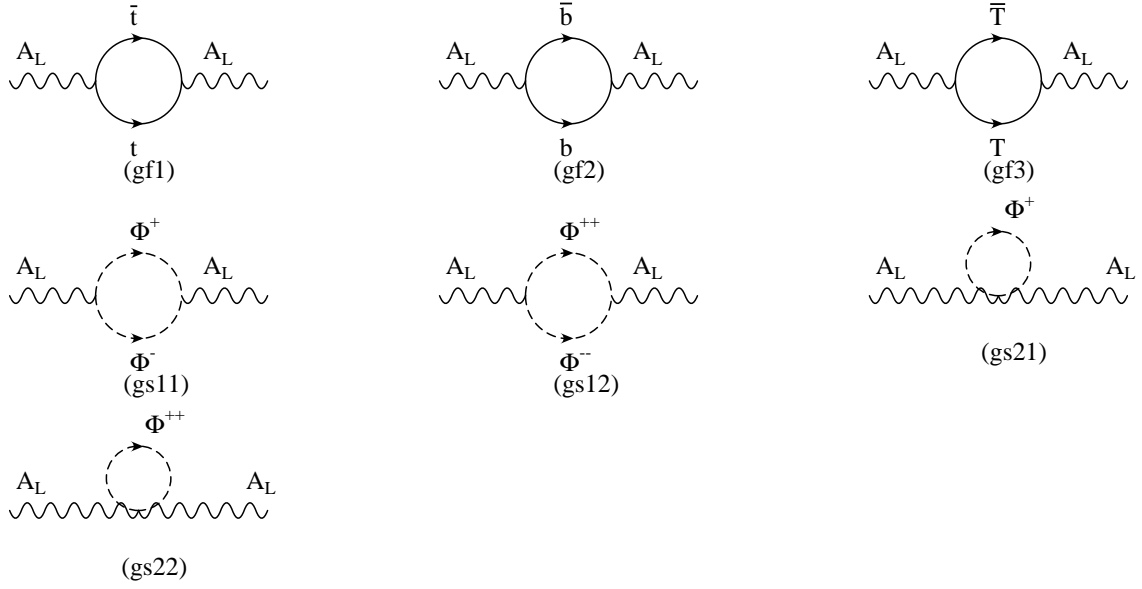}}
\caption{\label{fdpigg} Complete list of diagrams due to fermions and scalar fields 
to the self-energy of the photon, $A_{L}$. 
}
\end{figure}

\begin{figure}
{\center
\includegraphics[scale=0.75]{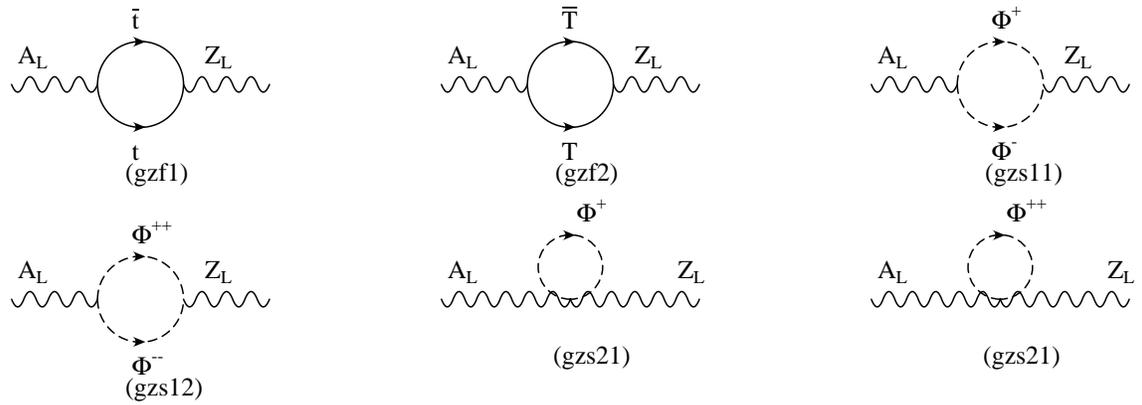}}
\caption{\label{fdpigz} Complete list of diagrams due to fermions and scalar fields 
to the self-energy $\Pi^{\gamma Z}$. 
}
\end{figure}

\begin{figure}
{\center
\includegraphics[scale=0.75]{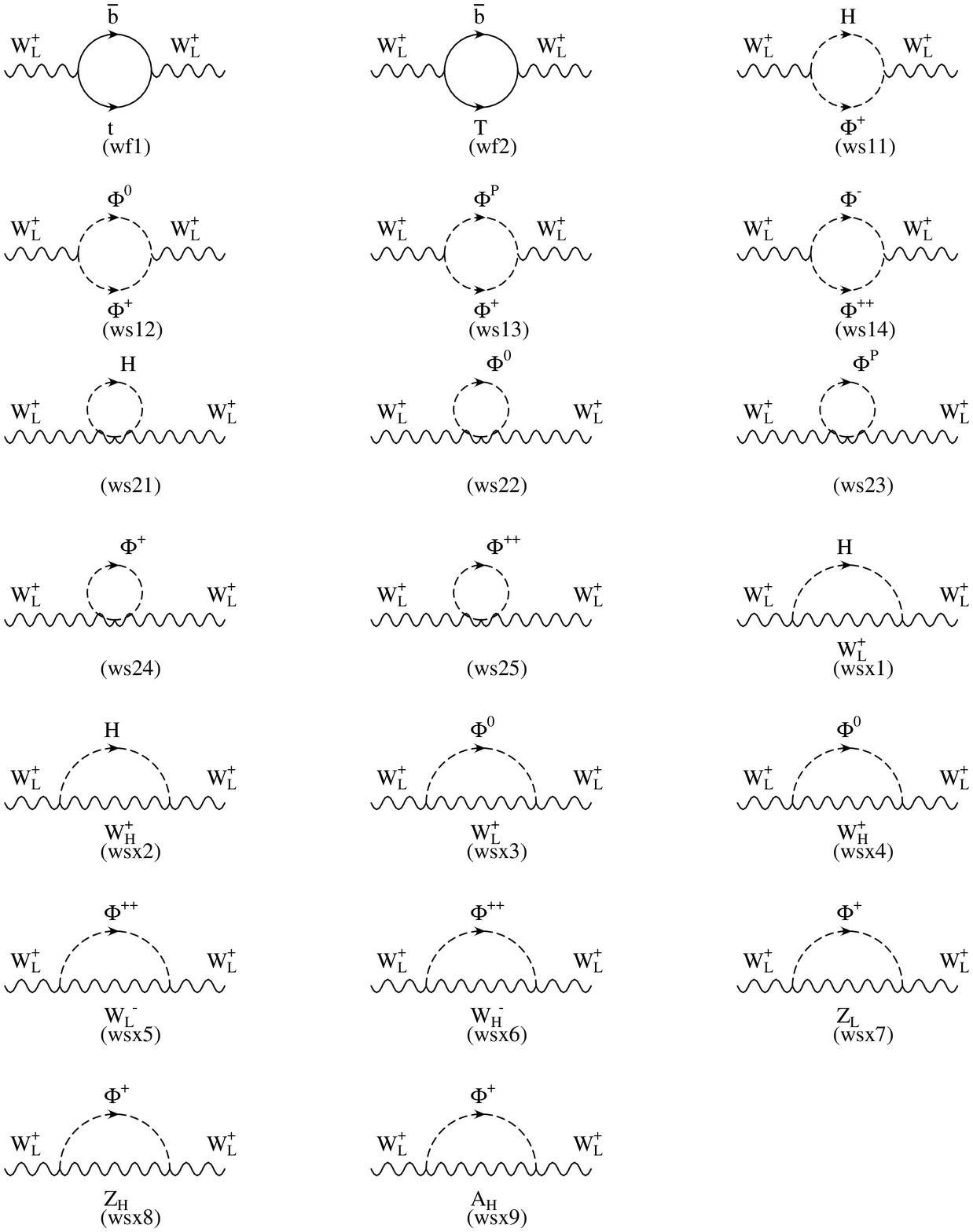}}
\caption{\label{fdpiww} Complete list of diagrams due to fermions and scalar fields 
to the self-energy of the Standard Model W gauge boson. 
}
\end{figure}

\begin{figure}
{\center
\includegraphics[scale=0.75]{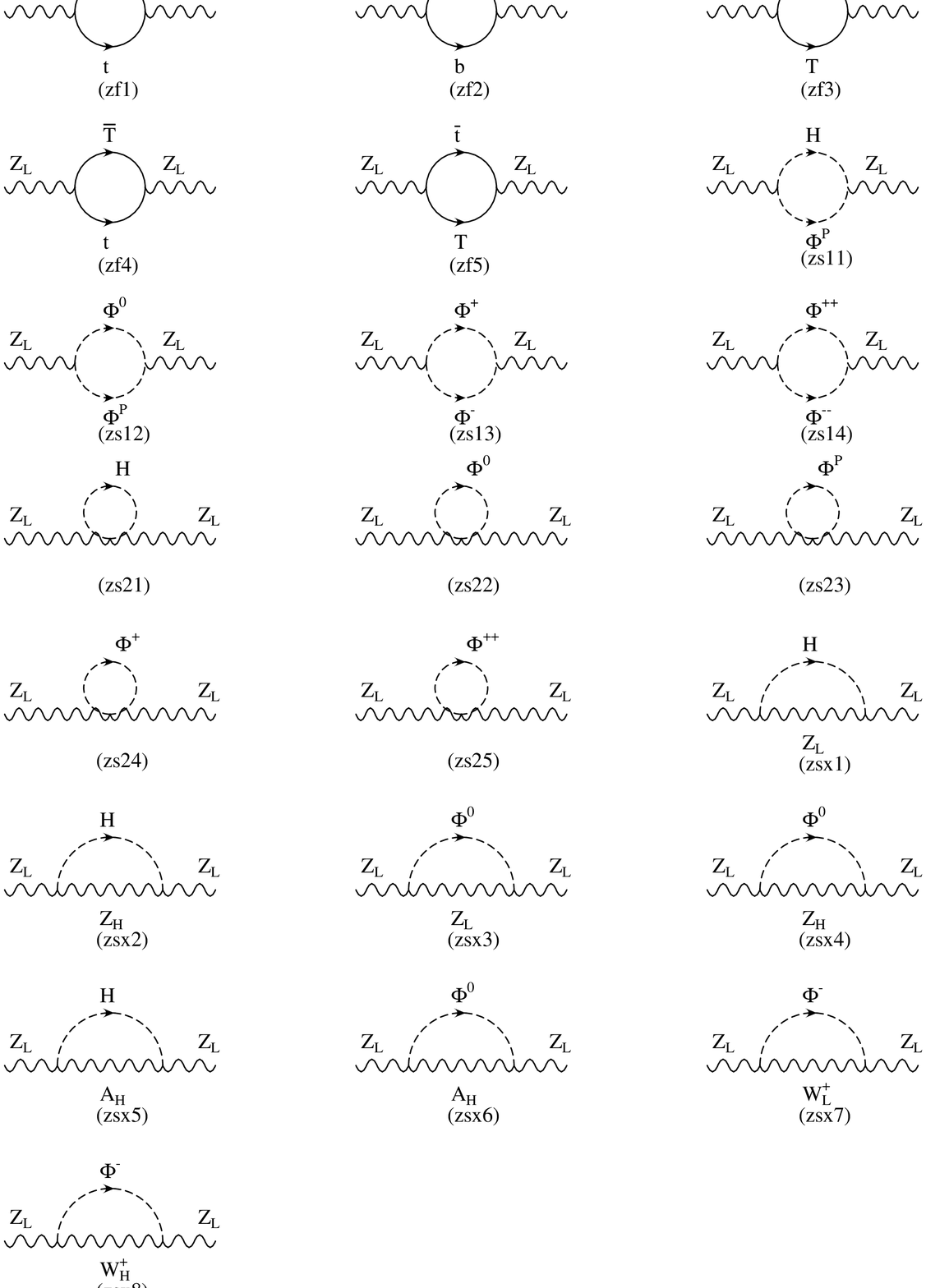}}
\caption{\label{fdpizz} Complete list of diagrams due to fermions and scalar fields 
to the self-energy of the Standard Model Z gauge boson. 
}
\end{figure}

\pagebreak

\bibliography{lh.v2}

\end{document}